\documentclass[aps,prb,twocolumn,superscriptaddress,10pt]{revtex4-2}

\usepackage{amsmath, amssymb}
\usepackage{mathrsfs}
\usepackage{graphicx}
\usepackage{bm}
\usepackage{braket}
\usepackage{here}
\usepackage[colorlinks=true, citecolor=blue]{hyperref}

\newcommand{\del}{\partial}
\newcommand{\diff}{\mathrm{d}}

\newcommand{\Hc}{\mathrm{H.c.}}

\newcommand{\imu}{\mathrm{i}}
\newcommand{\epn}{\mathrm{e}}

\newcommand{\ua}{\uparrow}
\newcommand{\da}{\downarrow}
\newcommand{\dg}{\dagger}
\newcommand{\la}{\langle}
\newcommand{\ra}{\rangle}
\newcommand{\al}{\alpha}
\newcommand{\be}{\beta}
\newcommand{\sg}{\sigma}

\newcommand{\gm}{\gamma}
\newcommand{\lam}{\lambda}

\newcommand{\T}{\mathrm{T}}
\newcommand{\dvec}[1]{\hspace{-1mm}\stackrel{\leftrightarrow}{#1}\hspace{-1mm}}
\newcommand{\nt}{\notag \\}

\newcounter{num}
\newcommand{\Rnum}[1]{\setcounter{num}{#1} \Roman{num}}

\newcommand{\mrm}[1]{\mathrm{#1}}
\newcommand{\mcal}[1]{\mathcal{#1}}
\newcommand{\mscr}[1]{\mathscr{#1}}

\begin{document}

\title{
Electron chirality and hydrodynamic helicity: Analysis in the atomic limit
}

\author{Tatsuya Miki}
\thanks{These authors contributed equally to this work.}
\affiliation{Institute for Materials Research (IMR), Tohoku University, Sendai, 980-8577, Japan}
\affiliation{Department of Physics, Saitama University, Sakura, Saitama 338-8570, Japan}
\affiliation{Department of Physics, Chiba University, Chiba 263-8522, Japan}
\author{Yuta Kakinuma}
\thanks{These authors contributed equally to this work.}
\affiliation{Department of Physics, Saitama University, Sakura, Saitama 338-8570, Japan}
\affiliation{Department of Physics, Chiba University, Chiba 263-8522, Japan}
\author{Masato Senami}
\affiliation{Department of Micro Engineering, Kyoto University, Kyoto 615-8540, Japan}
\author{\\Masahiro Fukuda}
\affiliation{Institute for Solid State Physics, The University of Tokyo, Chiba 277-8581, Japan}
\author{Michi-To Suzuki}
\affiliation{Department of Materials Science, Graduate School of Engineering, Osaka Metropolitan University, Sakai, Osaka 599-8531, Japan}
\affiliation{Center for Spintronics Research Network, Graduate School of Engineering Science, The University of Osaka, Toyonaka, Osaka 560-8531, Japan}
\author{Hiroaki Ikeda}
\affiliation{Department of Physical Sciences, Ritsumeikan University, Kusatsu, Shiga 525-8577, Japan}
\author{Shintaro Hoshino}
\affiliation{Department of Physics, Chiba University, Chiba 263-8522, Japan}
\affiliation{Department of Physics, Saitama University, Sakura, Saitama 338-8570, Japan}

\begin{abstract}
    Electron chirality has been proposed as a microscopic quantity that characterizes electronic handedness, yet its underlying control parameter has not been clearly identified. 
    Furthermore, its applicability is limited to systems with spin-orbit coupling, which motivates the need for alternative measures of chirality.
    In this work, we explore two complementary measures of chirality: electron chirality and hydrodynamic helicity.
    By analyzing a minimal atomic model under chiral crystal fields, we clarify how the interplay among crystal fields, spin-orbit coupling, and electron correlation gives rise to non-zero values of chirality measures.
    Although electron chirality increases with both spin-orbit coupling and chiral crystal field strength, the dependence on these two factors is highly non-trivial.
    Particularly, when the chiral crystal field is varied continuously and the energy levels approach quasidegenerate points, the electron chirality is insensitive to spin-orbit coupling, resulting in a remarkable enhancement of chirality.
    In contrast, the hydrodynamic helicity, defined as a two-body pseudoscalar quantity, remains non-zero even without spin-orbit coupling, originating from electron-electron interactions.
    Perturbative analysis reveals distinct symmetry selection rules governing the two quantities.
    Our results provide fundamental insight into the origin of chiralities in electronic systems.
\end{abstract}

\maketitle

\section{Introduction}

Among the various manifestations of asymmetry in nature, chirality, defined as non-superimposability with a mirror image, stands out as particularly significant.
Because of its fundamental nature, chirality is recognized as an important concept in various fields, including biology, chemistry, and particle physics.
Recently, in condensed matter physics, chiral materials have garnered growing attention due to their variety of unique phenomena, such as cross-correlation response, chiral magnetism \cite{Tokura18, Tokura21}, chirality-induced spin selectivity (CISS) \cite{Ray1999, Gohler2011, Inui2020, Ohe2024}, and circular dichroism \cite{Brinkman2024, Yen2024}.
These chiral phenomena have been observed in a wide range of materials, including organic systems \cite{Ray1999, Gohler2011} and inorganic crystals \cite{Inui2020, Ohe2024, Brinkman2024, Yen2024}.

Since electronic spin seems to connect with momentum in transport phenomena such as CISS and current-induced magnetization, one would expect that spin-orbit coupling (SOC) plays an important role in these phenomena.
Indeed, the importance of SOC has been experimentally pointed out in the CISS effect \cite{Adhikari2023}.
In contrast, chiral phenomena in materials do not necessarily require strong SOC.
For example, chiral phenomena can emerge even in the organic systems with intrinsically weak SOC \cite{Ray1999, Gohler2011}.

To understand and control the chiral phenomena in materials, the quantification of chirality is an important issue, beyond the binary distinction such as left- and right-handedness.
From the atomic positions point of view, molecular chirality can be described by a set of pseudoscalar parameters that quantify the degree of mirror-symmetry breaking \cite{Harris1997, Harris1999, Banik2016, Gomez-Ortiz2024, Bousquet2025}.
However, the pseudoscalar definition of chirality cannot be uniquely defined \cite{Harris1999, Banik2016}.
Alternatively, chirality can also be quantified in terms of the electrons that govern the material properties.
In the multipole classification, chirality can be quantified using a localized atomic basis \cite{Kugel1973, Ohkawa1983, Kuramoto2009, Santini2009}.
In particular, electric-toroidal multipoles capture the degree of mirror-symmetry breaking in the electronic structure \cite{Dubovik1986, Dubovik1990, Prosandeev2006, Guo2012, Hayami2019, Oiwa2022, Hayami2022, Hirose2022, Kishine2022, Hoshino2023, Kusunose2024, Inda2024, Hayami2024, Oiwa2025}, providing a framework to describe the behaviors of chiral materials \cite{Oiwa2022, Ishitobi25}.
In this context, a mechanism for controlling spin currents using junctions of chiral materials has also been proposed \cite{Matsubara2025}.

Such electronic asymmetries can be characterized through the spatial distribution of fundamental quantities defined locally in space.
The typical examples are the charge density and spin density, which characterize electric and magnetic properties such as electric polarization and magnetization.
Recently, electron chirality density \cite{Hara2012, Senami2019, Kuroda2022, Hoshino2023, Aucar2024, Miki2025}, which can be introduced by extending the framework beyond non-relativistic theory to relativistic quantum theory, has been discussed as a new fundamental quantity \cite{Hoshino2023}.
This quantity characterizes the chirality of electrons in materials.

The electron chirality has also been studied in quantum chemistry to describe the intrinsic handedness of electrons in molecular systems.
This quantity, when evaluated at atomic nuclei, has been shown to be closely related to the parity-violating energy difference between enantiomers \cite{Vester1959, Ulbricht1962, Szabo1999, Laerdahl2000, Bast2011, Hara2012, Senami2019, Kuroda2022}.
Furthermore, it has been reported that the spatial gradient of the electron chirality density induces an additional spin torque, suggesting that the spatial distribution of electron chirality plays an important role in spin-dependent phenomena \cite{Tachibana2012, Fukuda2016}.

In our previous work, we have quantitatively evaluated electron chirality using density functional calculations in solids \cite{Miki2025, Miki2025_julia}.
We compute the spatial distribution of electron chirality in chiral and axial crystals.
Furthermore, we proposed that electron chirality can be experimentally detected through circular dichroism in photoemission spectroscopy, which reflects the difference in response to right- and left-handed circularly polarized light.
The electron chirality provides a basis for further theoretical and experimental investigations of chiral electronic systems.

Although the electron chirality provides a measure of chirality, the underlying control parameter has not yet been clearly identified.
In particular, it is important to examine under what conditions, such as specific atomic configurations or charges in solids, electron chirality is enhanced.
Clarifying this control parameter is expected to result in the establishment of systematic guidelines for materials design and predictive principles for chiral electronic phenomena.
It is therefore necessary to capture the essential mechanisms for electron chirality as a first step.

Furthermore, since the electron chirality is derived from the relativistic quantum theory, it takes non-zero values only when we consider the relativistic effect in the Hamiltonian, namely in the presence of SOC.
This raises the question of whether another measure of the chirality of electrons can be defined for systems without SOC.
One possible candidate is obtained by extending our attention to the two-body quantity called hydrodynamic helicity, which was originally introduced in (classical) hydrodynamics \cite{Moffatt1992, Moffatt2014}.
By extending the concept of hydrodynamic helicity to the quantum electronic systems, we can consider the two-body chirality measure applicable to the system without SOC \cite{Hoshino2024}.

In this paper, we investigate the electron chirality and hydrodynamic helicity in the atomic limit and clarify how they arise from the interplay between chiral crystal fields, SOC, and electron interaction. 
Using a minimal atomic model with chiral crystal fields, we demonstrate that the electron chirality is significantly enhanced by tuning the chiral crystal fields and is insensitive to SOC near quasidegenerate energy levels.
Furthermore, we show that hydrodynamic helicity, defined as a two-body pseudoscalar quantity, arises from electron-electron interactions and remains non-zero even in the absence of SOC. 
These results clarify that the two chirality measures reflect distinct transformation properties under two-fold rotation and encode different physical aspects of electronic handedness.

This paper is organized as follows.
In Sec.~\ref{sec:def}, we define the two chirality measures, electron chirality and hydrodynamic helicity, and discuss their fundamental properties.
In Sec.~\ref{sec:model}, we introduce an atomic model with chiral crystal fields to investigate the microscopic origin of these quantities.
The numerical and perturbative results for the electron chirality are presented in Sec.~\ref{sec:chiral}, while Sec.~\ref{sec:hydro} discusses the hydrodynamic helicity in the interacting system.
Sec.~\ref{sec:summary} provides a summary and discussion.
The supplementary discussions are included in the Appendices.
Appendix~\ref{sec:berry} provides the Berry connection interpretation of electron chirality.
Appendix~\ref{sec:classical} discusses the classical case for the hydrodynamic helicity.
The additional results of the electron chirality, which focus on the second excited state, are presented in Appendix~\ref{sec:chirality_2nd}.
We present the numerical results for the hydrodynamic helicity in several cases: the same-orbital interaction in Appendix~\ref{sec:hydro_sameorb}, the effect of spin-orbit coupling in Appendix~\ref{sec:hydro_soc}, and the two-electron system in Appendix~\ref{sec:hydro_n2}.
Appendix~\ref{sec:momentum} summarizes helicity in momentum space as another chirality measure.

\section{Chirality densities \label{sec:def}} 

\subsection{Electron chirality \label{sec:chirality_def}}

The electron chirality is defined in terms of the Dirac field.
In the Weyl representation, the electron chirality is expressed as the difference between the right- and left-handed electron density \cite{Sakurai_book}:
\begin{align}
    \tau^Z(\bm r) = \psi_{\mrm R}^\dg(\bm r) \psi_{\mrm R}(\bm r) - \psi_{\mrm L}^\dg(\bm r) \psi_{\mrm L}(\bm r). \label{eq:chirality}
\end{align}
For low-energy physics, it is useful to employ the non-relativistic expression, which is represented by the Schr\"odinger field $\psi(\bm r) = (\psi_\ua(\bm r), \psi_\da(\bm r))^\T$ \cite{Berger2003, Fukuda2016, Hoshino2023, Hoshino2024, Miki2025}.
Performing $1/m$ expansion up to the leading order with electron mass $m$, we obtain the non-relativistic expression given by
\begin{align}
    \tau^Z(\bm r) = \frac{1}{2mc} \psi^\dg(\bm r) \dvec{\bm p}\cdot \bm \sg \psi(\bm r), \label{eq:chirality_nr}
\end{align}
where $c$ is the light velocity, $\bm \sg$ is the Pauli matrix, $\bm p = -\imu\hbar\nabla$ is the momentum operator, and we introduce the short-hand notation $A \dvec{O} B = A O B + (O^\ast A) B$.

Since electron chirality is derived from relativistic quantum theory, SOC is required for it to have non-zero expectation values.
To see this more explicitly, we consider the one-body part for the Hamiltonian $\mscr H = \int \diff\bm r \psi^\dg(\bm r) \mcal H(\bm r) \psi(\bm r)$ with $\mcal H(\bm r) = \bm p^2/2m + V(\bm r) -(\hbar e/4mc^2) \bm \sg \cdot (\bm E(\bm r) \times \bm p )$.
Here, the first term represents the kinetic energy, $V(\bm r)$ is a spin-independent potential, and the last term is the SOC.
In this case, we have the following commutation relation:
\begin{align}
    [\bm r \cdot \bm \sg, \mcal H(\bm r)] = \bigg(\frac{2m}{\imu\hbar}\bigg)\bm p \cdot \bm \sg + \frac{\imu\hbar e}{2mc^2}(\bm r \times \bm \sg) \cdot (\bm E(\bm r) \times \bm p). \label{eq:commut_chiral}
\end{align}
On the right-hand side, electron chirality appears in the first term, which arises from the commutator of $\bm r \cdot \bm \sigma$ with the kinetic energy part, while the second term arises from that with the SOC.
A similar relation in the relativistic case has also been considered \cite{Roberts2014, Hoshino2024}.
When we take the expectation value of an eigenstate, the left-hand side of Eq.~\eqref{eq:commut_chiral} is zero.
Therefore, the electron chirality vanishes in the absence of the second term, namely, without SOC.
Even without SOC, the off-diagonal matrix elements of the electron chirality do not necessarily vanish.
These matrix elements are, in fact, connected to the Berry-connection matrix with respect to the atomic positions of the nuclear potentials.
We provide a detailed explanation of this point in Appendix~\ref{sec:berry}.

We can further rewrite the expectation value of Eq.~\eqref{eq:commut_chiral} as
\begin{align}
    \la\bm p\cdot \bm \sg\ra
    &= - \frac{\hbar e}{2m^2c^2} \left\langle \bm E\cdot \left( \bm L\times \bm S - \frac{2m^2c^2}{e} \bm r\times \bm {\mathcal P}_S \right) \right\rangle,
    \label{eq:toroidal}
\end{align}
where we have defined the spin angular momentum $\bm S= \hbar \bm \sg/2$, orbital angular momentum $\bm L= \bm r\times \bm p$, and spin-derived electric polarization $\bm {\mathcal P}_S = (\hbar e/4m^2c^2)\bm p\times \bm\sg$ \cite{Wang2006, Hoshino2023}.
Here, the quantities such as $\bm L\times \bm S$ \cite{Wang2017, Chikano2021, Hayami2022, Hoshino2023, Inda2023} and $\bm r\times \bm {\mathcal P}_S$ \cite{Spaldin2008, Hayami2018, Kusunose2020, Bhowal2024} are known as the electric toroidal dipoles.
This expression provides another interpretation of the electron chirality: the value of $\la \tau^Z\ra$ in the stationary state is equivalent to the inner product of the electric field and electric toroidal dipole as defined by Eq.~\eqref{eq:toroidal}.
This can be regarded as a manifestation of the concept that the combination of polarity and axiality gives rise to chirality \cite{Hayami2025}, since the electric field and the electric toroidal dipole are polar (parity odd) and axial (parity even) vectors, respectively.

\subsection{Hydrodynamic helicity \label{sec:hydro_def}}

\subsubsection{Definition}

The electron chirality $\tau^Z(\bm r)$ takes non-zero values only in the presence of both SOC and chiral crystal fields.
This motivates us to consider whether there exist other chirality measures applicable to systems without SOC.
We show below that, by extending the notion of electron chirality from one-body to two-body quantities, one can construct such chirality measures even in the absence of SOC.

A representative example of such chirality measures is the hydrodynamic helicity $\bm v(\bm r) \cdot [\nabla \times \bm v(\bm r)]$, which characterizes the chirality of the velocity field $\bm v(\bm r)$ in fluid mechanics \cite{Moffatt1992, Moffatt2014}.
Here, we extend this by replacing the velocity field with the electronic current $\bm v(\bm r) \to \bm j(\bm r)$, which is given for electrons by
\begin{align}
    H(\bm r) = :\bm j(\bm r) \cdot [\nabla \times \bm j(\bm r)]:, \label{eq:hydro}
\end{align}
where $\bm j(\bm r) = \frac{e}{2m} \psi^\dg(\bm r) \dvec{\bm p} \psi(\bm r)$ is a current density with elementary charge $e$ \cite{Hoshino2024}.
The right-hand side of Eq.~\eqref{eq:hydro} includes the normal ordering ($:$) for the natural definition in quantum field theory. 
The physical or intuitive interpretation of $H(\bm r)$ can be provided for the classical case.
We provide the explanation in Appendix~\ref{sec:classical} from the perspective of analogy with Ginzburg-Landau theory, electromagnetic fields, and linking number.

We show that the electronic correlation is necessary in order to obtain the non-zero value of $\la H(\bm r) \ra$ in the system without SOC.
We expand the operator by a wave function whose quantum number is given by $\al$: $\psi_\sg(\bm r) = \sum_\al w_\al(\bm r) c_{\al\sg}$ and then 
\begin{align}
    \la H(\bm r) \ra &= - \frac{\hbar^2 e^2}{2m^2} \sum_{ijk\al\be\sg\sg'} \epsilon_{ijk}
    w_\al^\ast(\bm r) \partial_i w_\beta^\ast(\bm r) \partial_j w_\gm(\bm r) \partial_k w_\delta(\bm r) \nt
    &\times \la c_{\al\sg}^\dg c_{\beta\sg'}^\dg c_{\gm\sg'} c_{\delta\sg} \ra + \mrm{c.c.},
\end{align}
where we have introduced a short-hand notation $\del_i = \del / \del x_i$ and the anti-symmetric tensor $\epsilon_{ijk}$.
For a time-reversal symmetric case without SOC, $w_\al$ can be chosen as real.
Then, if the correlation is assumed to be absent, the helicity is zero by applying the Wick decomposition for the correlation function.
Hence, at least, the correlation effect is necessary for a non-zero value of $\la H(\bm r) \ra$.

The quantum hydrodynamic helicity can be non-zero even for a spinless fermion.
The spinless fermion is justified if we consider the Hamiltonian in the non-relativistic limit with the Zeeman effect.
Then, the spin-up and spin-down sectors are completely separated, and hence we only have to consider the spin-up sector, i.e., the spinless fermion model.
Note that the definition Eq.~\eqref{eq:hydro} can be applied to the spinful bosonic particles, while the helicity is zero for the spinless boson \cite{Hoshino2024}.
In the following, however, we assume the fermionic system since we focus on the electron chirality, which is important in condensed matter systems.

\subsubsection{Scalar spin chirality}

In a more general sense, for a given vector $\bm X(\bm r)$, the pseudoscalar quantity $\bm X(\bm r) \cdot [\nabla \times \bm X(\bm r)]$ serves as a measure of helicity or chirality. 
The quantity discussed above is distinctive in that the spin degree of freedom is irrelevant. 
In contrast, one may also introduce quantities that explicitly depend on spin. 
As an illustrative example, let us consider the following quantity:
\begin{align}
    \chi(\bm r,\bm r') &= \la :\bm s(\bm r) \cdot [\bm \nabla' \times \bm s(\bm r')]:\ra,
    \\
    \bm s(\bm r) &= \frac \hbar 2 \psi^\dg(\bm r) \bm \sg \psi (\bm r),
\end{align}
where $\bm s(\bm r)$ is the spin density operator.
The normal ordering is evaluated as
\begin{align}
    \chi(\bm r,\bm r') 
    &= \frac{\hbar}{2} \la \psi^\dg \dvec{\bm p}\cdot \bm \sg\psi\ra \delta(\bm r-\bm r') + \la \bm s(\bm r) \cdot [\bm \nabla' \times \bm s(\bm r')]\ra.
\end{align}
Hence, the one-body part is identified as the electron chirality density.
This gives a divergent contribution at $\bm r=\bm r'$, which is the reason for the necessity of the normal ordering.
Nevertheless, it is interesting to recognize that the electron chirality $\tau^Z(\bm r)$ is associated with the normal ordering of the two-body chirality measure.
Since we focus on the spin-independent definition of chirality, we consider the hydrodynamic helicity introduced in Eq.~\eqref{eq:hydro} below.

\section{Atomic model with crystal fields \label{sec:model}}

\begin{figure}[tb]
    \centering
    \includegraphics[width=8.5cm]{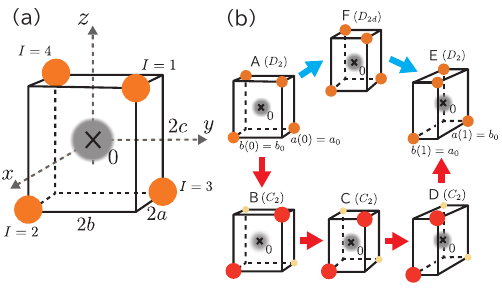}
    \caption{(a) Schematic figure for the atomic model. 
    (b) Configurations of crystal fields.
    The cross symbols indicate the nuclear potential at the origin, and the electronic cloud is illustrated as a gray shaded area. The red, orange, and yellow circles represent the charges of the crystal fields.}
    \label{fig:crystal_field}
\end{figure}

To see the origin of the electron chirality, we here consider the simplified atomic model with crystal fields, which is schematically shown in Fig.~\ref{fig:crystal_field} (a).
We consider the nuclear potential at the origin, represented by the cross symbol, and place the crystal fields at the four corners of a cube (shown as orange circles) whose edge lengths are $2a, 2b, 2c$.
The model Hamiltonian is given as follows:
\begin{align}
    \mscr H &= \mscr H_0 + \mscr H_{\mrm{CF}} + \mscr H_{\mrm{SOC}} +\mscr H_{\mrm{int}}, \\
    \mscr H_0 
    &= \int \diff\bm r\, \psi^\dg(\bm r) \bigg( \frac{\bm p^2}{2m} - \frac{Z_{\mrm n} e^2}{|\bm r|} \bigg) \psi(\bm r), \label{eq:h_0} \\
    \mscr H_{\mrm{CF}} &= \sum_{I=1}^N \int \diff\bm r\, \psi^\dg(\bm r) \frac{Q_I}{|\bm r - \bm R_I|} \psi(\bm r), \label{eq:h_cf} \\
    \mscr H_{\mrm{SOC}} &= -\frac{\hbar e}{4mc^2} \int \diff\bm r \psi^\dg(\bm r) \bm \sg \cdot \Big(-\frac{Z_{\mrm n} e \bm r}{r^3} \times \bm p \Big) \psi(\bm r), \label{eq:h_soc} \\
    \mscr H_{\mrm{int}} &= \frac{1}{2} \int \diff\bm r \diff\bm r' :\psi^\dg(\bm r) \psi(\bm r) \frac{e^2}{|\bm r - \bm r'|} \psi^\dg(\bm r') \psi(\bm r'):, \label{eq:h_int}
\end{align}
where $\mscr H_0$ represents the kinetic energy and nuclear potential at the origin with atomic number $Z_{\mrm n}$, $\mscr H_{\mrm{CF}}$ is the crystal field potential, $\mscr H_{\mrm{SOC}}$ is the SOC, and $\mscr H_{\mrm{int}}$ is the interaction among electrons.
In Eq.~\eqref{eq:h_cf}, $\bm R_I$, $Q_I$, and $N$, are the position, charge, and number of crystal fields, respectively.
As for SOC, the contribution from the crystal field decays as $|\bm r - \bm R_I|^{-2}$ in the limit where the positions of the crystal fields $\bm R_I$ are taken far from the origin.
Therefore, we consider only the SOC originating from the atom at the origin.
In the following, we set the atomic number as $Z_{\mrm n} = 4$, and numerical calculations are performed in atomic units.

In this paper, we perform the analysis under varying crystal-field configurations, as shown in Fig.~\ref{fig:crystal_field} (b).
The points A and E are the enantiomers connected by the spatial inversion combined with a rotation, and distinguished by the parameter $t$ ($t = 0$ and $t = 1$, respectively).
It is known that the two enantiomers can be connected either through an achiral point or along a path that does not pass through any achiral configuration in general \cite{Mezey1995, Weinberg1997, Harris1999}.
Following Ref.~\cite{Harris1999}, we consider two types of paths connecting points A and E: one passes through the achiral point F [A-F-E, blue arrows in Fig.~\ref{fig:crystal_field} (b)], while the other does not [A-B-C-D-E, red arrows in Fig.~\ref{fig:crystal_field} (b)], as shown in Fig.~\ref{fig:crystal_field} (b).
The positions of crystal field potentials $\bm R_I\, (I = 1,\cdots, 4)$ in Eq.~\eqref{eq:h_cf} are placed at 
\begin{align}
    \bm R_I(t) = 
    \begin{cases}
        (a(t), b(t), c(t)) \quad I = 1 \\
        (a(t), -b(t), -c(t)) \quad I = 2 \\
        (-a(t), b(t), -c(t)) \quad I = 3 \\
        (-a(t), -b(t), c(t)) \quad I = 4
    \end{cases}.
\end{align}
The color and size of the circles in Fig.~\ref{fig:crystal_field} (b) express the magnitude of charges for the crystal fields.

Along the path A-F-E, the edge lengths of the cube are parametrized as follows:
\begin{align}
    a(t) &= b_0 t + a_0(1 - t), \label{eq:cf_param_a} \\
    b(t) &= a_0 t + b_0(1 - t),  \label{eq:cf_param_b}\\
    c(t) &= 1.2, \label{eq:cf_param_c}
\end{align}
with $a_0 = 0.99, b_0 = 1.01$.
By varying $t$ as $0 \to 1$, the lengths of edges $a$ and $b$ are interchanged, thereby ensuring that A and E are related by spatial inversion.
Note that although $t$ can be extended to the regions $t < 0$ and $t > 1$, we focus on the region $0 \leq t \leq 1$ when we discuss the path in Fig.~\ref{fig:crystal_field} (b).
The achiral point F corresponds to $t = 0.5$, at which the two edge lengths are equal $a(t) = b(t)$.
The system has a two-fold rotational symmetry along $a$, $b$, and $c$ directions, and the point group is $D_2$ except for the F point whose point group is $D_{2d}$.
We fix the charges of the crystal fields as $Q_I = 5$ along this path.

On the other hand, along the path A-B-C-D-E, we first modify the charges of the crystal fields (A-B), then change the edge lengths according to Eqs.~\eqref{eq:cf_param_a}-\eqref{eq:cf_param_c} (B-C-D), and finally revert the charges to their original values (D-E).
Along the path A-B and D-E, we vary the charges of the crystal fields in Eq.~\eqref{eq:h_cf} according to the following parametrization:
\begin{align}
    Q_I(s, u) =
    \begin{cases}
    q_0 [1 + \mu (s + u)] \quad i = 1 \\
    q_0 [1 + \mu (s - u)] \quad i = 2 \\
    q_0 [1 - \mu (s + u)] \quad i = 3 \\
    q_0 [1 - \mu (s - u)] \quad i = 4
    \end{cases}, \label{eq:cf_param_qi}
\end{align}
with $q_0 = 5, \mu = 0.15$.
The parameter $s$ varies as $0 \to 1$ along the path A-B, and as $1 \to 0$ along D-E, and $u$ is fixed at $u = 0$.
Note that $Q_I$ along the path A-F-E corresponds to $(s, u)= (0, 0)$.
Since the charges of the crystal field change, the system is chiral throughout the path A-B-C-D-E, and the two-fold rotational symmetry along only the $a$-axis is preserved. 
Thus, the point group of the system is $C_2$.
By connecting these two paths as A-B-C-D-E-F-A, a periodic path can be formed.
In the following sections, the calculations are performed along this path.
Note that we use the parameter $u$ to investigate the case of $C_2$-symmetry breaking configurations in Secs.~\ref{sec:chiral_c2} and \ref{sec:hydro_c2}.

\begin{figure}[tb]
    \centering
    \includegraphics[width=8.5cm]{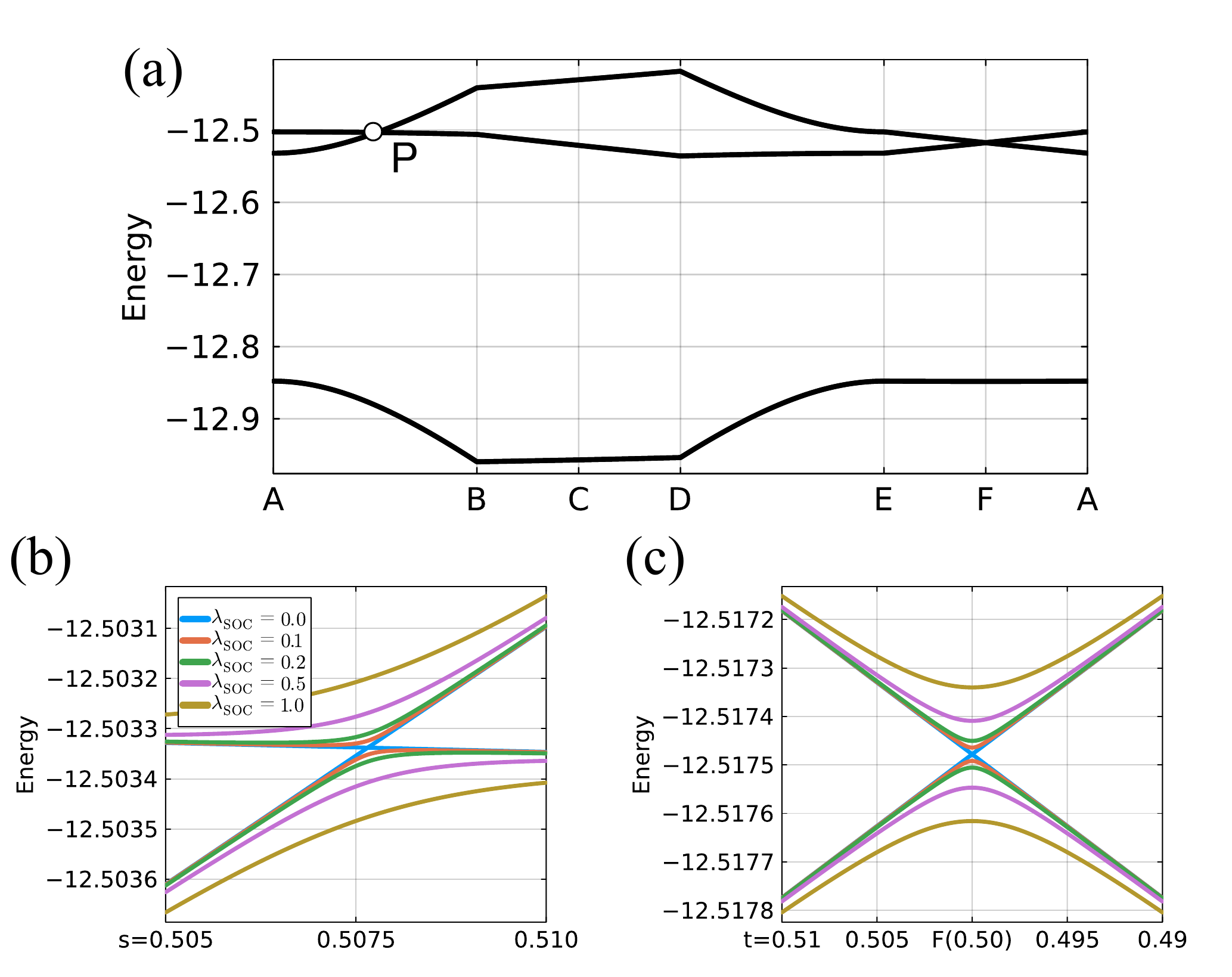}
    \caption{(a) Energy diagram of the lowest three eigenstates. 
    (b) Enlarged view of energy diagram of the first and second excited state near the point P $[(t, s) = (0.0,0.505) \to (0.0,0.51)]$ for several values of $\lambda_{\mrm{SOC}}$.
    (c) Enlarged view of energy diagram of the first and second excited state near the point F $[(t, s) = (0.51, 0.0) \to (0.49, 0.0)]$.
    }
    \label{fig:energy}
\end{figure}

Hereafter, we focus on the localized electronic state of the atom at the origin.
Namely, we expand the field operator by the eigenfunction of $\mscr H_0$ as $\psi(\bm r) = \sum_{nlm} R_{nl}(r) Y_{lm}(\hat{\bm r}) c_{nlm\sg}$, where $R_{nl}(r)$ is a radial part and $Y_{lm}(\hat{\bm r})$ is the spherical harmonics.
Then, we make it possible to deal with a finite-dimensional Hilbert space by restricting the space to a few selected orbitals.
For the non-zero electron chirality, parity mixing is required. 
In addition, we have numerically confirmed that the electron chirality is vanishingly small when we consider only $s$-$p$ orbital mixing in this setup.
Thus, in this paper, we consider $\{2p, 3d\}$ orbitals as a minimal set.

\section{Results for electron chirality \label{sec:chiral}}

\begin{figure}[tb]
    \centering
    \includegraphics[width=7.0cm]{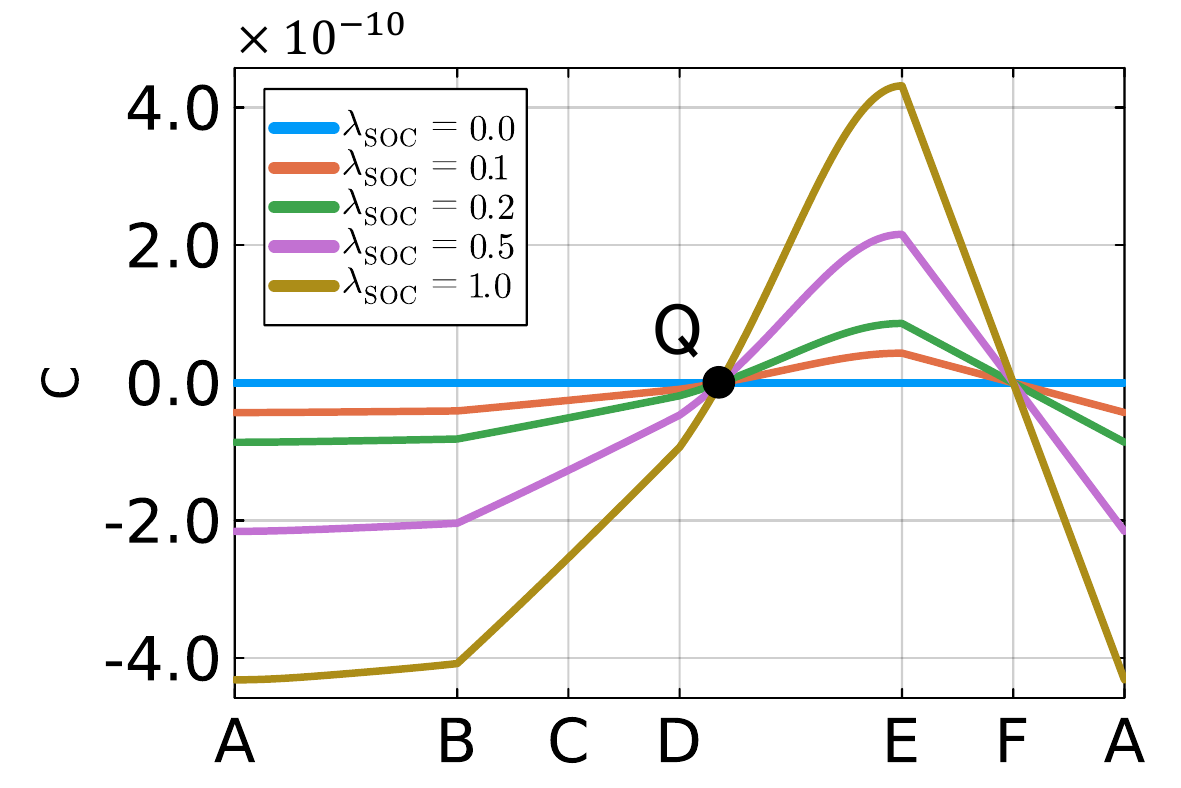}
    \caption{Electron chirality in the ground state. 
    The point where the electron chirality vanishes is denoted as Q.
    }
    \label{fig:chirality_gs}
\end{figure}

In this section, we investigate the behavior of the electron chirality.
Specifically, we focus on the total chirality defined by
\begin{align}
    C = \frac{1}{N_{\mrm{degen}}^{\mcal P}} \sum_{n\in \mcal P} \int\diff\bm r\, \braket{n |\tau^Z(\bm r)| n}, \label{eq:chirality_tot}
\end{align}
where $\mcal P$ is the subspace of interest, and $N_{\mrm{degen}}^{\mcal P}$ is the degeneracy of $\mcal P$.
We consider two low-energy cases for $\mcal P$: the ground state and the first excited state.
Since the low-energy states have $p$-like character (including the ground, first excited, and second excited states), we present the results for the electron chirality of the second excited state in Appendix~\ref{sec:chirality_2nd} for completeness.
Note that a non-zero value of Eq.~\eqref{eq:chirality_tot} arises only in the presence of SOC as discussed in Sec.~\ref{sec:chirality_def}.
Below, we clarify the dependence of electron chirality on two key ingredients: the chiral crystal field and the SOC.

\subsection{One-electron system: Numerical results}

\begin{figure*}[tb]
    \centering    
    \includegraphics[width=18.0cm]{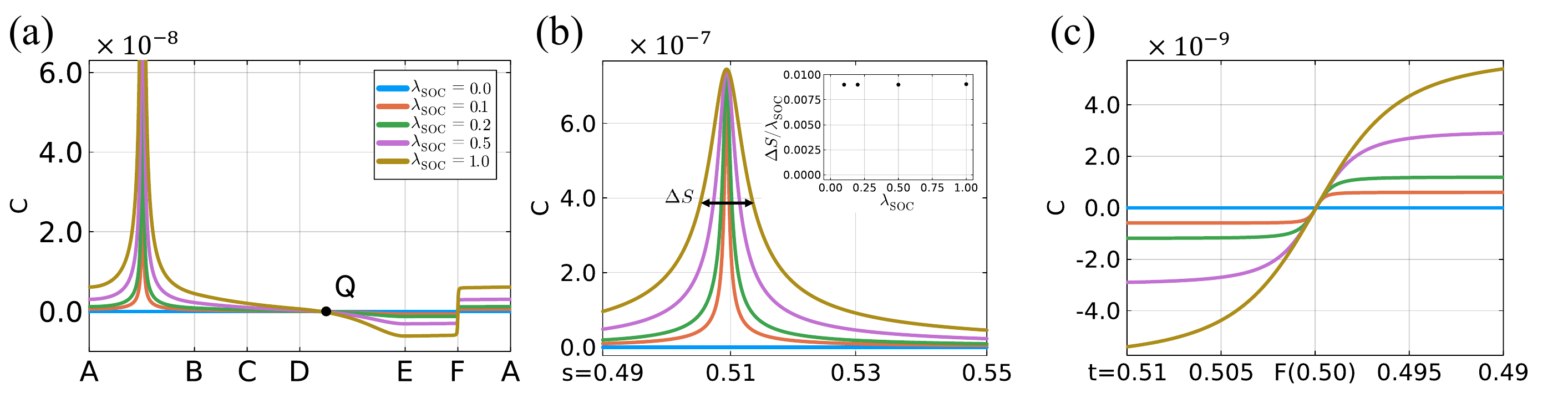}
    \caption{
    (a) Electron chirality in the first excited state.  
    (b) Enlarged view of the electron chirality of the first excited state near the peak [$(t,s)=(0.0,0.49)\rightarrow(0.0,0.55)$].
    The inset shows the distribution of the full width at half maximum $\Delta S$ for each SOC strength. 
    (c) Enlarged view of the first excited-state electron chirality near the F point [$(t,s)=(0.51,0.0)\rightarrow(0.49,0.0)$].
    }
    \label{fig:chirality_1st}
\end{figure*}

We begin by considering the one-particle energies. 
To examine the effect of SOC, we introduce a parameter $\lambda_{\mathrm{SOC}}$ that controls its strength and consider the Hamiltonian $\mathscr{H} = \mathscr{H}_0 + \mathscr{H}_{\mathrm{CF}} + \lambda_{\mathrm{SOC}} \mathscr{H}_{\mathrm{SOC}}$.  Figure~\ref{fig:energy} (a) shows the three lowest one-particle energy levels for $\lam_{\mrm{SOC}} = 1$ along the crystal-field configuration A-B-C-D-E-F-A in Fig.~\ref{fig:crystal_field} (b). Each state is two-fold degenerate because of time-reversal symmetry, and the dominant components of the corresponding wave functions are derived from the $p$ orbitals. 
Note that the five energy bands of $d$-orbital character appear at higher energy (about $4\ \mathrm{Hartree}$ above).
Except at the F point, inversion symmetry is broken at all chiral crystal field configuration points, which gives rise to mixing between $p$ and $d$ orbitals (parity mixing).
In the first and second excited states, a quasi-degeneracy occurs at the F point and at a point along the A$\to$B path, which we denote as P [more precisely defined by the peak of the electron chirality, see Fig.~\ref{fig:chirality_1st} (b)].
At point P, a tiny gap is opened by SOC as shown in Fig.~\ref{fig:energy} (b), thereby the bands are not four-fold degenerate for $\lam_{\mrm{SOC}} > 0$.
The energy splitting for $\lam_{\mrm{SOC}} = 1$ is the order of $10^{-3}$.
A similar quasi-degeneracy is also present at point F, as shown in Fig.~\ref{fig:energy} (c).
These quasi-degeneracies play a crucial role in the anomalous behavior of electron chirality as shown below.

First, we consider the case where $\mcal P$ is set to the ground state.
Figure~\ref{fig:chirality_gs} shows the total electron chirality for several values of $\lam_{\mrm{SOC}}$.  
At the achiral point (F), the total electron chirality vanishes owing to mirror symmetry.  
As the chiral crystal field configuration deviates from the F point, the electron experiences a chiral crystal field, leading to an increase in $C$.

It is also noteworthy that there exists a zero crossing of $C$ between the crystal-field configurations D and E.
This arises because, in association with the sign change at the achiral point, $C$ must necessarily cross zero along the path.
In other words, since the electron chirality changes sign from positive to negative along the E$\to$F$\to$A sequence, it must cross zero from negative to positive somewhere along the A$\to$E path. 
We refer to such an accidental zero-crossing point as Q in the following, which is distinct from the symmetry-protected zero point F.
We also plot $C$ for the several SOC strength $\lam_{\mrm{SOC}}$ in Fig.~\ref{fig:chirality_gs}.
The electron chirality increases almost linearly with $\lam_{\mrm{SOC}}$.

Next, we consider the case where $\mcal P$ is the first excited state. 
Physically, this situation corresponds to a three-electron configuration in the weak-interaction limit.  
In this case, the two lowest one-particle levels are fully occupied, and their contributions to the electron chirality are almost negligible; thus, the dominant contribution arises from the partially filled first excited state [see also the ground state of three-electron calculation in Fig.~\ref{fig:chirality_int} (b)].  
In contrast to the ground state, the excited state has quasi-degenerate energy at P and F.

Figure~\ref{fig:chirality_1st} (a) shows the electron chirality, which exhibits characteristic behavior in the vicinity of the quasi-degenerate points. 
The electron chirality $C$ changes rapidly near the point F, while its magnitude is markedly enhanced in the vicinity of the P point.

To investigate the origin of these behaviors, we further examine the dependence on the SOC strength $\lambda_{\rm SOC}$.
Here, we focus on the characteristic points P and F, for which the origin of the behaviors will be discussed in Sec.~\ref{sec:chirality_perturb}.
As shown in Fig.~\ref{fig:chirality_1st} (b), the peak value of the electron chirality at the P point is almost independent of the SOC. 
Furthermore, as shown in the inset of Fig.~\ref{fig:chirality_1st} (b), which presents the full width at half maximum $\Delta S$, the peak width increases approximately in proportion to the strength of the SOC.
The enlarged view of the region near the F point is shown in Fig.~\ref{fig:chirality_1st} (c).  
When approaching the F point, one finds that the behavior of the electron chirality is essentially independent of the SOC strength.  
Thus, as the system approaches a quasi-degenerate point, the electron chirality becomes insensitive to the magnitude of the SOC.

To see the behavior of the peak at point P in more detail, we consider the electron chirality on the $t\mathchar`-s$ plane.
Figure~\ref{fig:ts-plane} (a) shows the distribution of the electron chirality on the $t\mathchar`-s$ plane on a symmetric logarithmic scale, where the ridge structure and zero points are seen along the P-line (green line) and Q-line (orange line), respectively.
When we consider a loop with $\mrm{max}\, t = 1 - \mrm{min}\, t$ indicated by the black arrow, the existence of point P is guaranteed by the $C_2$ symmetry, as discussed in Sec.~\ref{sec:chiral_c2}.
On the other hand, the existence of point Q is ensured by the fact that this loop goes through two enantiomers at $(t, s) = (\mrm{max}\, t, 0)$ and $(t, s) = (\mrm{min}\, t, 0)$.
Figure~\ref{fig:ts-plane} (b) shows the peak height at the point P for electron chirality, which increases with increasing distance from the achiral F point (characterized by the length $L$).
This increasing behavior can be explained from the perturbation analysis in Sec.~\ref{sec:chirality_perturb}.

\subsection{Perturbative interpretation \label{sec:chirality_perturb}}

\subsubsection{Energy scales in the atomic model}

We now attempt to provide a qualitative understanding of the above non-trivial behavior.
Let us begin with a summary of the energy scales inherent in the present system.
The largest scale is the energy separation between the $p$ and $d$ orbitals, denoted by $E_{dp}$. 
A deviation from the F point breaks inversion symmetry, and the associated energy scale is denoted by $E_{\text{chiral-CF}}$.  
The crystal-field splitting unrelated to chirality is denoted by $E_{\mathrm{CF}}$, which is roughly given by the energy difference between the first and second excited states.  
There is also the energy scale associated with the SOC, $E_{\mathrm{SOC}}$.
In general, these scales satisfy $E_{dp} \gg E_{\mathrm{CF}} > E_{\text{chiral-CF}}, E_{\mathrm{SOC}}$, and, except near the achiral F point, one typically has $E_{\text{chiral-CF}} \gg E_{\mathrm{SOC}}$.

\begin{figure}[tb]
    \centering
    \includegraphics[width=8.5cm]{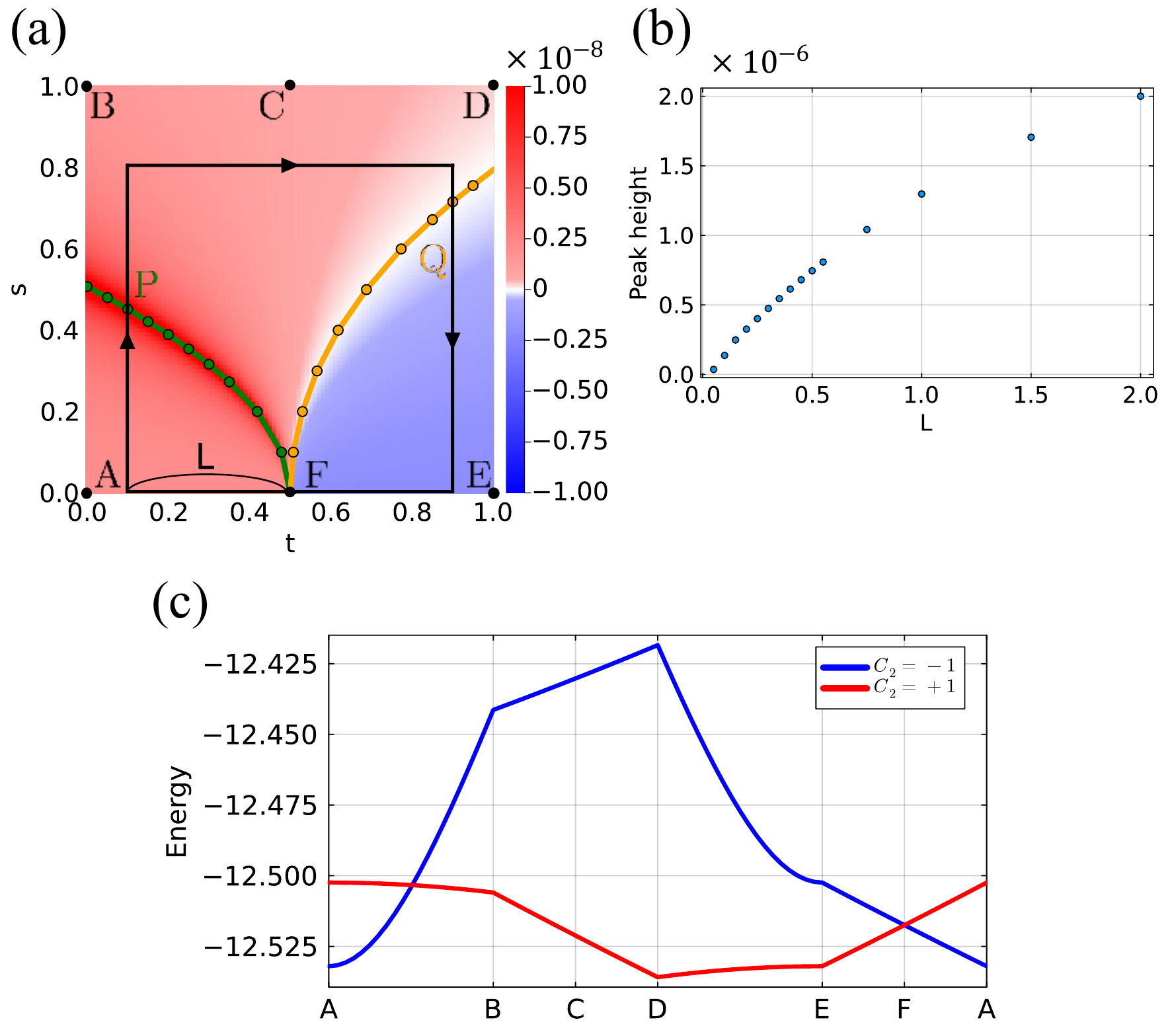}
    \caption{
    (a) Positions of the peak points P (green line) and zero points Q (orange line) on the $t\mathchar`-s$ plane of electron chirality in a symmetric logarithmic scale.
    (b) $L$-dependence of peak height at point P for electron chirality.
    (c) $C_2$ eigenvalues for first and second excited states.
    }
    \label{fig:ts-plane}
\end{figure}

A rough estimate from the calculated energy diagram gives  
$E_{dp} \sim 4$, $E_{\mathrm{CF}} \sim 0.5$, $E_{\text{chiral-CF}} \sim 0.1$, and $E_{\mathrm{SOC}} \sim 0.001$.  
Note that $E_{\text{chiral-CF}}$ represents the typical magnitude of the chiral crystal field, and it vanishes at the F point.
It should be noted that these energy scales do not necessarily correspond to the splitting of eigenenergies, but rather to the typical magnitudes of the terms appearing in the Hamiltonian.  
For example, at the P point, the energy scale of the chiral crystal field, $E_{\text{chiral-CF}}$, is non-zero, while the actual energy splitting ($\ll E_{\text{chiral-CF}}$) is nearly zero.

\subsubsection{First-order perturbation theory}

With the energy scales in mind, we now discuss the electron chirality using perturbation theory with respect to the SOC.  
Using Brillouin-Wigner perturbation theory~\cite{Kuramoto_book}, the electron chirality, to first order in the SOC, is given by
\begin{align}
    C &\simeq
    \frac{1}{N_{\mrm{degen}}^{\mcal P}}
    \sum_{n \in \mathcal P_0} 
    \int \diff \bm r \nt 
    &\quad\times\braket{ n| \tau^Z(\bm r) \ (E_n -\mathscr H^{(0)})^{-1} \mathscr Q_0 \mathscr H_{\mrm{SOC}} |n }_0
    + {\rm c.c.}. \label{eq:chirality_pertub}
\end{align}
Here, $\mathcal{P}_0$ denotes the non-perturbed subspace of interest,  
$\mscr H^{(0)} = \mscr H_0 + \mscr H_{\mrm{CF}}$ is the unperturbed Hamiltonian, and $\mscr Q_0$ is the projection operator onto the complement states of $\mcal P_0$.  
Hereafter, the subspace projected by $\mscr Q_0$ is denoted as $\mcal Q_0$.

For simplicity, we restrict $\mathcal{Q}_0$ to include only the single non-perturbed state that is expected to contribute most strongly, i.e., the one closest in energy to the $\mathcal{P}_0$ subspace,  
so that the total number of states considered amounts to four.
Noting that we are considering one-particle wave functions and upon tracing out the spin degrees of freedom, we obtain
\begin{align}
    C &\simeq
    \frac{2}{mc N_{\mrm{degen}}^{\mcal P}} \sum_{\alpha \in \mathcal P_0} \sum_{\beta \in \mathcal Q_0} \nt
    &\quad\times\frac{\langle \alpha | \frac{\hbar Z_{\mrm n}e^2}{4m^2c^2 r^3} \bm L | \beta \rangle_0 \cdot \langle \beta | \bm p | \alpha \rangle_0 }{E_\alpha - E^{(0)}_\beta}
    +{\rm c.c.}, \label{eq:chirality_pertub_2}
\end{align}  
where $\bm L = \bm r\times \bm p$, 
$|\alpha\rangle_0, |\beta\rangle_0$ denote one-particle states, and $E_\alpha$ and $E_\beta^{(0)}$ are the eigenenergies of $\mathscr H$ and $\mathscr H^{(0)}$, respectively.  
In the numerator, the energy scale $E_{\mathrm{SOC}}$ appears along with the matrix element $\langle \alpha | {\bm p} | \beta \rangle$.  
The latter, $\langle \alpha | {\bm p} | \beta \rangle = - i \hbar \langle \alpha | \nabla | \beta \rangle$, arises from parity mixing due to the breaking of spatial inversion symmetry, and is therefore of the order of $\hbar E_{\text{chiral-CF}} / (a_0 E_{dp})$, where $a_0$ is the Bohr radius.
Near the degeneracy point, the denominator in Eq.~\eqref{eq:chirality_pertub_2} is expected to be small, leading to an enhancement of $C$ \cite{Berger2003}.

\subsubsection{Application to the ground states}

First, when choosing the ground state as $\mathcal{P}$,  
the energy denominator is of the order of $E_{\mathrm{CF}}$  
(here, $|\alpha\rangle$ refers to the ground state and $|\beta\rangle$ to the first excited state).  
Accordingly, the order-of-magnitude estimate for the electron chirality is  
\begin{align}
C \sim \frac{\hbar}{m c a_0} \cdot \frac{E_{\mathrm{SOC}}}{E_{\mathrm{CF}}} \cdot \frac{E_{\text{chiral-CF}}}{E_{dp}}, \label{eq:kaki1}
\end{align}
which implies a natural result: the electron chirality depends linearly on both the chiral crystal field and the SOC.

\subsubsection{Application to the first excited states}

Next, when choosing the first excited state as $\mathcal{P}$, let us consider a generic point other than the F or P points (here, $|\alpha\rangle$ denotes the first excited state, and $|\beta\rangle$ is taken as the second excited state, which contributes most significantly).  
The energy appearing in the denominator is of the order of $E_{\text{chiral-CF}}$.  
This cancels with the same factor in the numerator, giving
\begin{align}
C \sim \frac{\hbar}{m c a_0} \cdot \frac{E_{\mathrm{SOC}}}{E_{\text{chiral-CF}}} \cdot \frac{E_{\text{chiral-CF}}}{E_{dp}} 
\ \propto \  \frac{E_{\mathrm{SOC}}}{E_{dp}}. \label{eq:chiral_gs}
\end{align}
Hence, the electron chirality is independent of the magnitude of the chiral crystal field,  
which explains the jump-like behavior near the F point.  
Here, it should be noted that we have assumed $E_{\text{chiral-CF}} \gg E_{\mathrm{SOC}}$.

If one approaches the F or P point closely enough, the energy denominator becomes of the order of the SOC, and one obtains  
\begin{align}
C \sim \frac{\hbar}{m c a_0} \cdot \frac{E_{\mathrm{SOC}}}{E_{\mathrm{SOC}}} \cdot \frac{E_{\text{chiral-CF}}}{E_{dp}} 
\ \propto \  \frac{E_{\text{chiral-CF}}}{E_{dp}}, \label{eq:per_near_F,P}
\end{align}
which is independent of the SOC but still linearly proportional to the chiral crystal field.  
Hence, as seen in Fig.~\ref{fig:chirality_1st} (b), at the quasi-degenerate P point, the electron chirality is insensitive to the SOC, while moving away from P, it decreases forming the peak structure.  
Similarly, in the immediate vicinity of the quasi-degenerate F point [immediate vicinity of F in Fig.~\ref{fig:chirality_1st} (c)], $C$ is independent of the SOC and grows linearly with the chiral crystal field away from F, which is consistent with Fig.~\ref{fig:ts-plane} (b).

From Eq.~\eqref{eq:chirality_pertub_2}, it is clear that the presence of quasi-degeneracy in the chiral region leads to an enhancement of the electron chirality.  
A natural question, however, is why quasi-degenerate points appear in the chiral region in the first place.
The degeneracy at achiral points is naturally understood as being protected by symmetry, but the existence of degeneracy in the chiral region may seem counterintuitive.  
In fact, in the absence of SOC and under $C_2$ symmetry, one can show that degeneracies at the achiral point and the one in the chiral region always appear in pairs.  
All the crystal fields considered in this paper respect $C_2$ symmetry along the $x$ axis.  
Thus, we can choose the energy eigenstate as a simultaneous eigenstate of the two-fold rotation operator along the $x$ axis:
\begin{align}
    R(C_2) &= \exp\bigg[-\imu\frac{\pi}{\hbar} L_z \bigg], \label{eq:c2}
\end{align}
with $L_z = -\imu \hbar \int \diff \bm r \psi^\dg (y\del_z - z\del_y) \psi$.
Considering a loop that includes an achiral point like F, the exchange of eigenvalues $C_2=\pm 1$ requires an even number of inversions as seen in Fig.~\ref{fig:ts-plane} (c).
Therefore, a degeneracy in the chiral region necessarily accompanies that at the achiral point.  
Such degeneracies are analogous to those discussed in the Nielsen-Ninomiya theorem in (1+1)d~\cite{Nielsen1981} and in symmetry-protected nodal-line semimetals~\cite{Kawakami2017}.  
When a small SOC is introduced in this situation, it leads to quasi-degeneracy and consequently enhances the electron chirality.
In this way, the peak of chirality is associated with the energy-crossing point in systems with chiral crystal fields without SOC.

Although the occurrence of the level crossing is guaranteed for the reasons discussed above, its position along the A--B path segment is accidental.
Therefore, upon changing the system parameters, the location of the crossing shifts.
A similar argument applies to the chirality zeros, namely the Q line in Fig.~\ref{fig:ts-plane} (a), whose positions depend sensitively on the system parameters.
Note that both the P and Q lines are connected to the achiral F point.

Let us comment on the relationship between enantiomeric configurations and $C_2$ symmetry. 
As discussed below, the $C_2$ eigenvalues are the convenient label for distinguishing the two enantiomers A and E for the first excited state.
These two enantiomeric configurations are related by a spatial inversion followed by a four-fold rotation around the $z$ axis ($IC_4$).
For electronic states with $p_x$ or $p_y$ character, which form the first and second excited states, since the $IC_4$ operation maps one enantiomer onto the other while reversing the sign of $p_x$ and $p_y$ components under $C_2$, the two configurations have opposite $C_2$ eigenvalues.
Therefore, the $C_2$ eigenvalue provides a label for distinguishing the enantiomers in the first excited state.

\subsection{Electron chirality without \texorpdfstring{$C_2$}{C2}-symmetry \label{sec:chiral_c2}}

\begin{figure}[tb]
    \centering    \includegraphics[width=8.5cm]{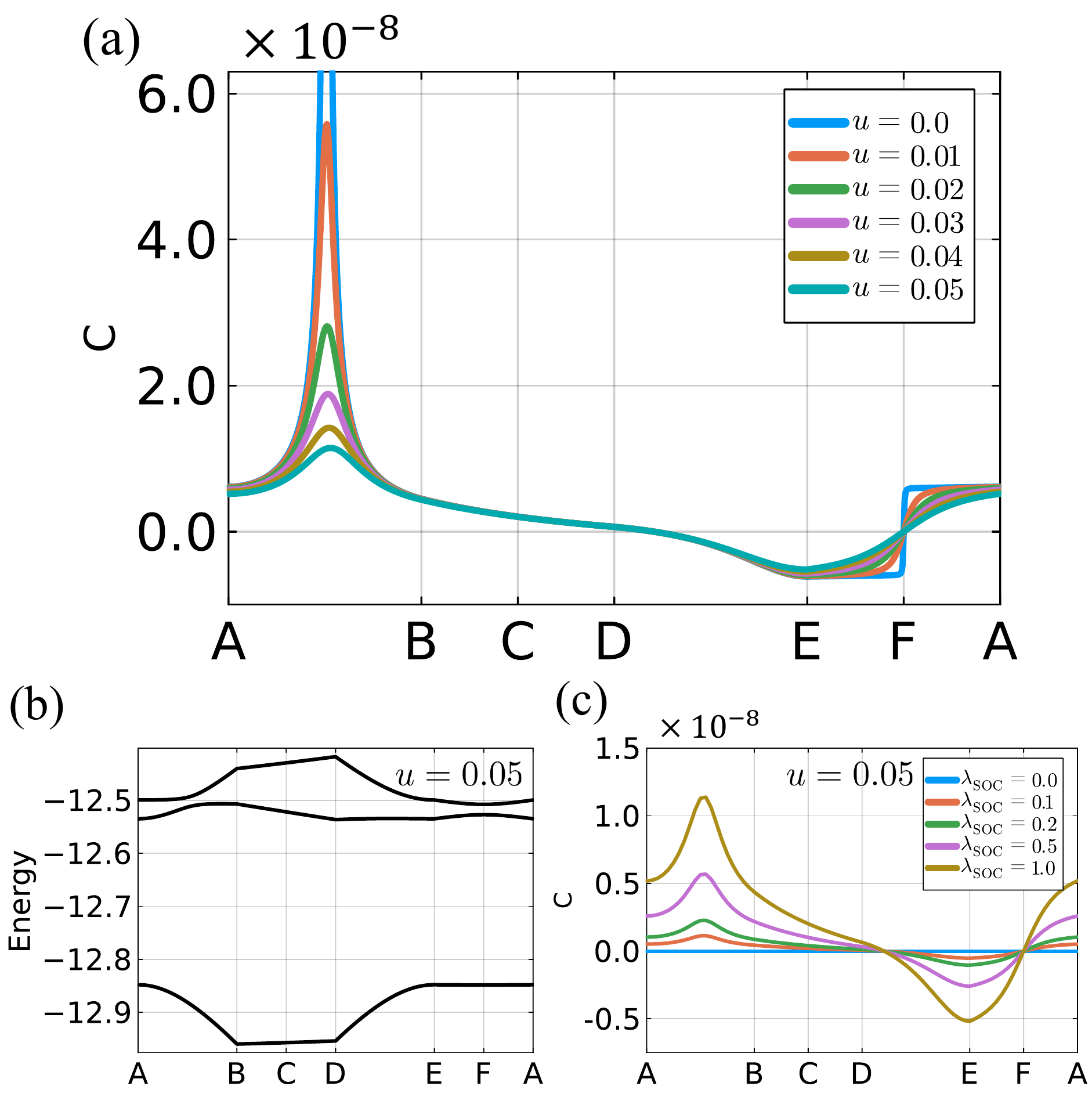}
    \caption{
    (a) Electron chirality of the first excited state for $u \in [0.0, 0.05]$, which breaks $C_2$ symmetry.  
    (b) Energy bands for $u=0.05$.  
    (c) Electron chirality for $u=0.05$ as the SOC strength $\lambda_{\mathrm{SOC}}$ is varied over $[0,1]$.
    }
    \label{fig:chirality_u}
\end{figure}

In Sec.~\ref{sec:chirality_perturb}, we discussed the case in which $C_2$ symmetry is preserved.  
Here, we consider how the peak at the P point and the jump-like behavior near the F point change when $C_2$ symmetry is broken.  
By taking the parameter $u$ in Eq.~(\ref{eq:cf_param_qi}) to be non-zero, the point-group symmetry is reduced to $C_1$,  
and the electron chirality of the first excited states is shown in Fig.~\ref{fig:chirality_u} (a).  
The peak at the P point is lowered, and the discontinuous behavior near the F point is more gradual.  
This can be attributed to the lifting of the quasi-degeneracy at the P and F points due to the breaking of the $C_2$ symmetry of the crystal field, as illustrated in Fig.~\ref{fig:chirality_u} (b).  

Considering again the system to first order in perturbation theory, the disappearance of quasi-degeneracy increases the energy differences so that the energy denominators in the electron chirality near the P and F points are now determined by the crystal-field-induced splitting (parameterized by $u$).  
As a result, a discussion analogous to that for the ground state applies, and the electron chirality is proportional to both the chiral crystal field and the SOC, as in Eq.~(\ref{eq:kaki1}).  
Indeed, plotting the SOC dependence as in Fig.~\ref{fig:chirality_u} (c) shows that the electron chirality near the P and F points now varies with the SOC.

\subsection{Electronic interaction effect}

\begin{figure}[tb]
    \centering
    \includegraphics[width=7.7cm]{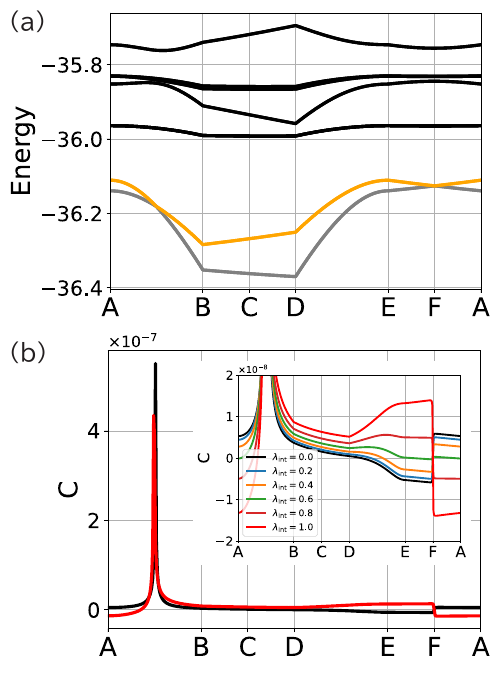}
    \caption{(a) Energy diagram and (b) ground state electron chirality with (red) and without (black) electron-electron interaction for the three-electron case. The inset in (b) shows an enlarged view of electron chirality around $C = 0$ for several values of $\lambda_{\mrm{SOC}}$.}
    \label{fig:chirality_int}
\end{figure}

We consider the electron-electron interaction effect on the electron chirality.
The corresponding Hamiltonian is given as $\mscr H = \mscr H_0 + \mscr H_{\mrm{CF}} + \mscr H_{\mrm{SOC}} + \lam_{\mrm{int}}\mscr H_{\mrm{int}}$, where we can control the magnitude of interaction by changing $\lam_{\mrm{int}}$.
To compare with the first excited state of the one-electron case, we here consider the three electrons on the $p$ and $d$ orbitals in the system with the same crystal fields as in Fig.~\ref{fig:crystal_field}.
The energy diagram for $\lam_{\mrm{int}} = 1$ is shown in Fig.~\ref{fig:chirality_int} (a).
Each state is doubly degenerate because of the time reversal symmetry.
The lowest two states (gray and orange lines) exhibit energy crossings between points A and B and at the achiral point F, in correspondence with those found in the first and second excited state of the one-electron system.

Figure~\ref{fig:chirality_int} (b) shows the ground-state electron chirality for the non-interacting ($\lambda_{\mathrm{int}} = 0$) and interacting ($\lambda_{\mathrm{int}} = 1$) cases.
We also present the $\lam_{\mrm{int}}$-dependence on the electron chirality in the inset.
The peak and sudden sign change are maintained even in the interacting case (red line).
However, as shown in the inset of Fig.~\ref{fig:chirality_int} (b), quantitative differences can be observed; for instance, the electron chirality changes sign around points A and F.

This change due to the electron-electron interaction is analogous to the variation of crystal field parameters, indicating that it can be understood within the one-body approximation, such as the Hartree approximation.

\section{Results for hydrodynamic helicity \label{sec:hydro}}

\subsection{Numerical results \label{sec:hydro_result}}

We discuss the many-body measure of chirality in the absence of SOC.
This type of chirality is characterized by the hydrodynamic helicity defined in Eq.~\eqref{eq:hydro}.
In the following, we consider a Hamiltonian without SOC but including electron-electron interactions, given by 
\begin{align}
    \mscr H = \mscr H_0 + \mscr H_{\mrm{CF}} + \lambda_{\mrm{int}}\mscr H_{\mrm{int}},
\end{align}
where the strength of the interaction is controlled by a parameter $\lambda_{\rm int} \in [0,1]$.
The positions and charges of the crystal fields are parametrized by Eqs.~\eqref{eq:cf_param_a}-\eqref{eq:cf_param_qi}.
The SOC effect for the hydrodynamic helicity is discussed in Appendix~\ref{sec:hydro_soc}.
We investigate the total helicity for the ground state defined by 
\begin{align}
    Z = \frac{1}{N_{\mrm{degen}}^{\mcal P}} \sum_{n\in \mcal P} \int\diff\bm r\, \braket{n |H(\bm r)| n}, \label{eq:hydro_tot}
\end{align}
where $\mcal P$ is the subspace of the degenerate ground states.

\begin{figure}[tb]
    \centering
    \includegraphics[width=8.5cm]{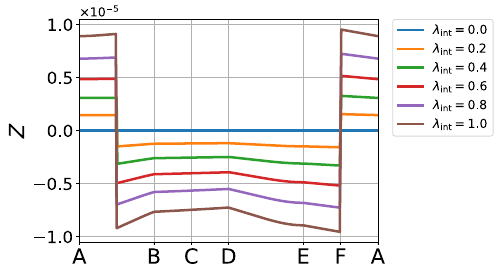}
    \caption{Hydrodynamic helicity of the ground state for several values of $\lambda_{\mrm{int}}$.}
    \label{fig:hydro}
\end{figure}

Figure~\ref{fig:hydro} shows the $\lambda_{\mrm{int}}$-dependence of hydrodynamic helicity for the ground state.
In the non-interacting case $\lam_{\mrm{int}} = 0$, the hydrodynamic helicity is zero, which is consistent with the discussion in Sec.~\ref{sec:hydro_def}.
Once the interaction is introduced ($\lambda_{\mrm{int}} > 0$), the hydrodynamic helicity increases linearly with the interaction strength.
In contrast to the behavior of electron chirality, the hydrodynamic helicity exhibits a rapid sign change even in the chiral crystal fields (A-B), corresponding to the energy crossing point in Fig.~\ref{fig:chirality_int} (a).
Apart from these two sign-changing points, the value remains almost constant with only minor variations.

Note that we have checked that the hydrodynamic helicity is almost reproduced by including only the same-orbital interaction, indicating that the local contribution is dominant in this model (see Appendix~\ref{sec:hydro_sameorb}).

\subsection{Perturbation analysis \label{sec:perturbation_hydro}}

In Sec.~\ref{sec:chirality_perturb}, we demonstrated that the peak in the electron chirality originates from an energy crossing. 
In contrast, as shown in Fig.~\ref{fig:hydro}, no such peak arises in the case of hydrodynamic helicity $Z$ in Eq.~\eqref{eq:hydro_tot}.
In the following, we examine the underlying reason for this difference from the standpoint of perturbation theory.

Since $Z$ increases linearly with respect to $\lambda_{\mrm{int}}$ in Fig.~\ref{fig:hydro}, we evaluate $Z$ within the first-order perturbation of the electron-electron interaction $\mscr H_{\mrm{int}}$:
\begin{align}
    Z &\simeq \frac{1}{N_{\mrm{degen}}^{\mcal P}} \sum_{n \in \mathcal P_0} \sum_{m \in \mathcal Q_0} \nt
    &\quad\times\frac{\int \diff \bm r \braket{n| H(\bm r) | m}_0 \braket{m | \mathscr H_{\mrm{int}} |n}_0}{E_n^{(0)} - E_m^{(0)}} 
    + {\rm c.c.},
    \label{eq:hydro_perturb}
\end{align}
where $\mcal P_0$ is the non-perturbed subspace of interest, and $\mcal Q_0$ is the complement subspace of $\mcal P_0$ [see also Eqs.~\eqref{eq:chirality_pertub} and \eqref{eq:chirality_pertub_2}].
where $\braket{n|\cdots|m}_0$ represents the matrix element between the eigenstates of the non-interacting Hamiltonian, and $E_{n}^{(0)}$ is a corresponding eigenenergy.

According to Eq.~\eqref{eq:hydro_perturb}, one might expect that the factor $1/(E_n^{(0)} - E_m^{(0)})$ would diverge at the degeneracy point, thereby suggesting an enhancement analogous to that of electron chirality.
However, as shown in Fig.~\ref{fig:hydro}, no such divergence actually occurs.
This is because the matrix elements of both the interaction part of the Hamiltonian $\braket{m | \mscr H_{\mrm{int}} | n}_0$ and hydrodynamic helicity $\int \diff\bm r\braket{n | H(\bm r) | m}_0$ vanish, which is guaranteed by the selection rule of $C_2$ rotational symmetry in group theory as discussed below.

As discussed in Sec.~\ref{sec:model}, the $C_2$ symmetry is preserved throughout the path.
Thus, we can choose $\ket{n}_0$ and $\ket{m}_0$ as eigenstates of the $C_2$ operator $R(C_2)$ in Eq.~\eqref{eq:c2}.
If we do not take spin rotation into account, we have $R(C_2)\ket{n} = \eta_n \ket{n}$, where $\eta_n = \pm 1$ is the eigenvalue of $R(C_2)$.
Since $\mscr H_{\mrm{int}}$ is symmetric under $C_2$, we obtain the following relation:
\begin{align}
    \braket{m|\mscr H_{\mrm{int}}|n}_0 = \eta_m \eta_n \braket{m|\mscr H_{\mrm{int}}|n}_0. \label{eq:selection}
\end{align}
Then, the matrix element $\braket{m|\mscr H_{\mrm{int}}|n}_0$ is zero for $\eta_m \eta_n = -1$.
Indeed, $\eta_n$ of the ground state and first excited state have opposite signs.
Note that the matrix element of hydrodynamic helicity $\int\diff\bm r \braket{n | H(\bm r) | m}_0$ satisfies a similar selection rule.
Therefore, $Z$ in Eq.~\eqref{eq:hydro_perturb} does not enhance even in the vicinity of the degeneracy points.

Here, we have chosen the non-interacting states $\ket{n}_0$ as the basis.
However, since the $C_2$ symmetry is preserved even in the presence of interactions, the same argument applies to the interacting eigenstates $\ket{n}$. In other words, the selection rule in Eq.~\eqref{eq:selection} remains valid upon replacing $\ket{n}_0$ with $\ket{n}$. This implies that, for the two low-energy states (gray and orange lines in Fig.~\ref{fig:chirality_int}), the hydrodynamic helicity $\int\diff\bm r \braket{n | H(\bm r) | m}_0$ is diagonal.

For the electron chirality $C$, the spin rotation must be taken into account for considering the selection rule of the matrix elements of $C$ and $\mscr H_{\mrm{SOC}}$.
Therefore, the selection rule Eq.~\eqref{eq:selection} does not hold.
In this way, the electron chirality and hydrodynamic helicity behave differently due to the intrinsic symmetry.

\begin{figure}[tb]
    \centering
    \includegraphics[width=8.5cm]{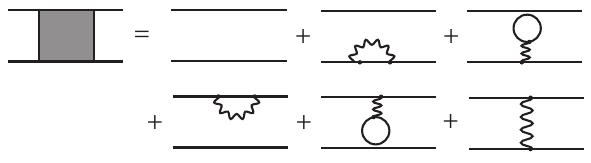}
    \caption{Feynman diagrams for the first-order perturbation of the electron-electron interaction.
    The straight lines represent electron propagators, and the wavy lines represent interactions among electrons.
    }
    \label{fig:diagram}
\end{figure}

A different perspective, based on Green’s function methods, leads to a deeper understanding of the above perturbation analysis.
We here employ the basis of the eigenfunction of the non-interacting Hamiltonian $\mscr H_0 + \mscr H_{\mrm{CF}}$.
The two-particle Green's function is defined by
\begin{align}
    &G^{\mrm{\Rnum{2}}}_{n_1 n_2 n_3 n_4 ; \sg_1 \sg_2 \sg_3 \sg_4}(\tau_1 \tau_2 \tau_3 \tau_4) \nt 
    &= \la \mcal T c_{n_1 \sg_1}(\tau_1) c_{n_2 \sg_2}(\tau_2) c_{n_3 \sg_3}^\dg(\tau_3) c_{n_4 \sg_4}^\dg(\tau_4) \ra, \label{eq:gfunc}
\end{align}
where $\mcal{T}$ represents imaginary time ordering and $O(\tau) = \epn^{\tau \mscr{H}} O \epn^{-\tau \mscr{H}}$ is the Heisenberg representation with imaginary time.
We now express the expectation value of hydrodynamic helicity $\la H(\bm r) \ra$ in terms of the two-particle Green's function in Eq.~\eqref{eq:gfunc}.
The field operator is expanded in the eigenfunctions of $\mscr H^{(0)}$ as
$\psi_s(\bm r) = \sum_n \phi_n^s(\bm r)c_n$, and the hydrodynamic helicity can be written as
\begin{align}
    \la H(\bm r) \ra &= -\frac{\hbar^2 e^2}{2m^2} \sum_{ijk} \sum_{n_1 \cdots n_4} \sum_{ss'} \nt &\times\lim_{\tau_1 \to \tau_2+0^+} \lim_{\tau_2 \to \tau_3+0^+} \lim_{\tau_3 \to \tau_4+0^+} \lim_{\tau_4 \to 0} \nt 
    &\times \epsilon_{ijk} \phi_{n_1}^{s\ast}(\bm r) \del_j \phi_{n_2}^{s'\ast}(\bm r) \del_k \phi_{n_3}^{s'}(\bm r) \del_i \phi_{n_4}^s(\bm r) \nt 
    &\times G^{\mrm{\Rnum{2}}}_{n_4 n_3 n_2 n_1 ; \sg_4 \sg_3 \sg_2 \sg_1}(\tau_4 \tau_3 \tau_2 \tau_1) + \Hc
\end{align}
Within the first-order perturbation of the electron interaction, the contributed diagrams are shown in Fig.~\ref{fig:diagram}.

Let us comment on the calculation of hydrodynamic helicity in solids.
While the evaluation based on the wave function should be difficult, the diagrammatic approach can evaluate the chiralities in this paper in the weak-interaction cases.
Since the hydrodynamic helicity is zero for the one-body approximation and linearly increases with respect to the electron interaction, as shown in Fig.~\ref{fig:hydro}, the ladder-type diagram (the last term in Fig.~\ref{fig:diagram}) should be important.

\subsection{Effect of \texorpdfstring{$C_2$}{C2} symmetry breaking \label{sec:hydro_c2}}

Now, we examine the behavior of the hydrodynamic helicity $Z$ defined in Eq.~\eqref{eq:hydro_tot} under $C_2$-symmetry breaking.
A central and unexpected finding of this analysis is a strong correlation between the hydrodynamic helicity $Z$ and the expectation value of the $C_2$ rotational operator defined in Eq.~\eqref{eq:c2}.
This correlation persists robustly under various parameter conditions.

\begin{figure}[tb]
    \centering
    \includegraphics[width=7cm]{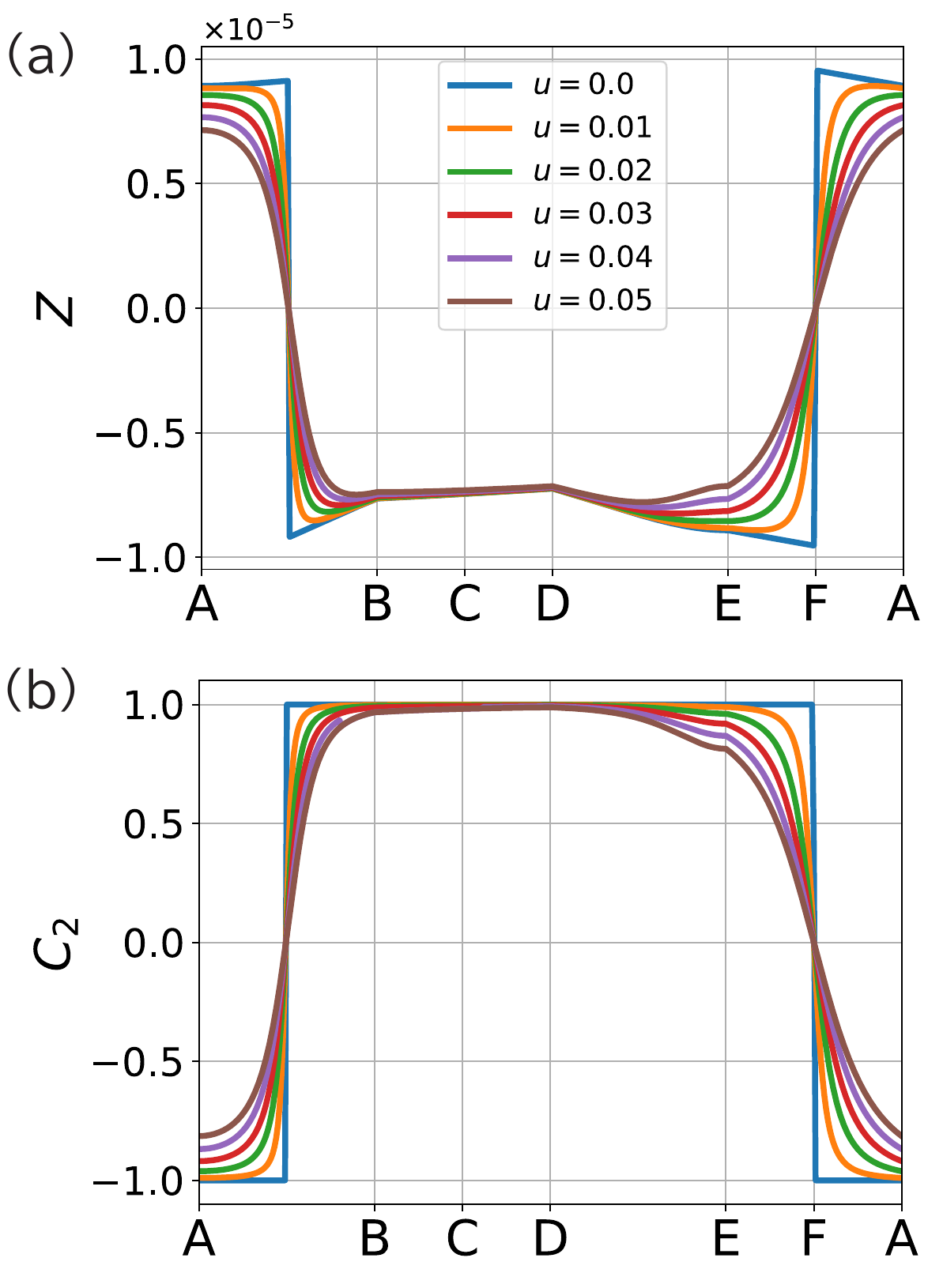}
    \caption{Effect of $C_2$-symmetry breaking and SOC for hydrodynamic helicity. (a) hydrodynamic helicity and (b) expectation value of $C_2$ operator for several values of $u$ (without SOC).}
    \label{fig:hydro_u}
\end{figure}

For the $C_2$-symmetry breaking case, we use the parameter $u$ in Eq.~\eqref{eq:cf_param_qi}.
The $C_2$ symmetry is breaking at $u > 0$.
Figure~\ref{fig:hydro_u} (a) shows the hydrodynamic helicity for several $u$.
As discussed in Sec.~\ref{sec:hydro_result}, $Z$ has sudden sign changes for the $u=0$ case.
However, as $u$ increases, this sign change is gradually broadened.

To gain further insight, we analyze the expectation value of the $C_2$ rotational operator for several $u$, whose result is shown in Fig.~\ref{fig:hydro_u} (b).
We can clearly see the correlation between $Z$ in Fig.~\ref{fig:hydro_u} (a) (with a minus sign) and the expectation value of the $C_2$ rotational operator in Fig.~\ref{fig:hydro_u} (b).
A similar correlation can be seen even in the system with SOC as shown in Appendix~\ref{sec:hydro_soc}.

Finally, we clarify the reason for this correlation between the hydrodynamic helicity and the expectation value of $C_2$ operator by focusing on two low-energy states.
Here, we take into account the two lowest-energy states: the ground state and the first excited state.
The key point is that, at the $C_2$ symmetric case [$u=0$ in Eq.~\eqref{eq:cf_param_qi}], both the $C_2$ operator and the hydrodynamic helicity are diagonalized by the low-energy eigenstates, the latter of which follows from the selection rule derived in Sec.~\ref{sec:perturbation_hydro}.
Namely, within the low-energy subspace of the Hilbert space, the $C_2$ operator and the hydrodynamic helicity commute.
In the following, we consider only one state of each Kramers pair for simplicity, since the time reversal symmetry is preserved.
At $u = 0$, we label the energy eigenstates by the eigenvalue of the $C_2$ operator as $R(C_2) \ket{\psi_\pm} = \pm \ket{\psi_\pm}$.
When $u$ is turned on ($u > 0$), the ground state $\ket{\psi}$ is given by the superposition of $\ket{\psi_+}$ and $\ket{\psi_-}$, and we represent it as $\ket{\psi} = U_+ \ket{\psi_+} + U_- \ket{\psi_-}$.
Since $\ket{\psi_+}$ and $\ket{\psi_-}$ diagonalize the hydrodynamic helicity in the low-energy subspace, we have $\int\diff\bm r\, H(\bm r) \ket{\psi_\pm} \simeq Z_\pm \ket{\psi_\pm}$, where $Z_\pm$ are corresponding expectation values for $\ket{\psi_\pm}$.
The expectation values for the ground state $\ket{\psi}$ are given by
\begin{align}
    Z &= \int\diff\bm r\, \braket{\psi | H(\bm r) | \psi} = |U_+|^2 Z_+ + |U_-|^2 Z_-, \label{eq:z_uv} \\
    C_2 &= \braket{\psi | R(C_2) | \psi} = |U_+|^2 - |U_-|^2. \label{eq:c2_uv}
\end{align}
From Eqs.~\eqref{eq:z_uv} and \eqref{eq:c2_uv}, we have the relation:
\begin{align}
    Z = \frac{Z_+ + Z_-}{2} + \frac{Z_+ - Z_-}{2} C_2. \label{eq:z_c2}
\end{align}
Thus, the hydrodynamic helicity $Z$ is a linear function of $C_2$ at low energy.

\begin{figure}[tb]
    \centering
    \includegraphics[width=6.5cm]{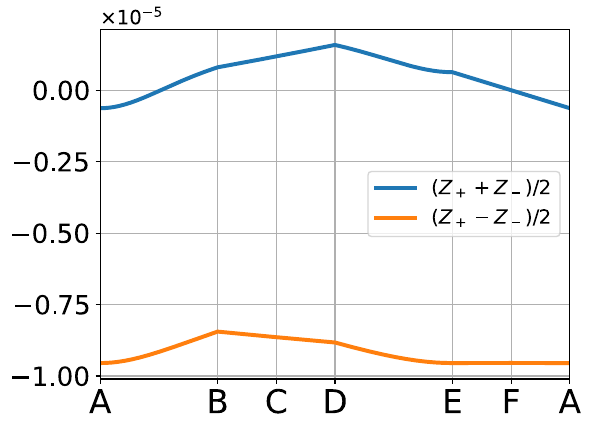}
    \caption{
    The values of $(Z_+ + Z_-)/2$ and $(Z_+ - Z_-)/2$.
    Since the two degenerate eigenstates have the same value of $Z_\pm$, we choose the one eigenvalue for hydrodynamic helicity for the plot.
    }
    \label{fig:z_c2}
\end{figure}

In particular, at the achiral point F, the two states $\ket{\psi_+}$ and $\ket{\psi_-}$ are degenerate.
Then, the states can be chosen such that the mirror reflection $\sg_d$, which is the symmetry operation at the F point, acts as $R(\sg_d) \ket{\psi_\pm} = \ket{\psi_\mp}$ up to an overall phase.
Since hydrodynamic helicity $\int \diff\bm r H(\bm r)$ is a pseudoscalar quantity, namely an odd function under the mirror reflection $\sg_d$, we obtain the relation 
\begin{align}
    Z_+ 
    &= \int \diff\bm r \braket{\psi_+ | H(\bm r) | \psi_+} \nt 
    &= -\int \diff\bm r \braket{\psi_- | H(\bm r) | \psi_-} = -Z_-.
\end{align}
Therefore, the first term in Eq.~\eqref{eq:z_c2} is zero at the F point.
Figure~\ref{fig:z_c2} shows the behavior of $(Z_+ \pm Z_-)/2$ for the ground state and first excited state.
One finds that $(Z_+ \pm Z_-)/2$ is insensitive to the variation of the crystal field.
Since the first term in Eq.~\eqref{eq:z_c2}, $(Z_+ + Z_-)/2$, is zero at the point F, it varies slightly around zero at the other points.
Thus, hydrodynamic helicity $Z$ and the expectation value of the $C_2$ operator exhibit similar behaviors, and the rapid sign change in $Z$ originates from the behavior of the $C_2$ operator.
This result provides valuable insight: the behavior of $Z$ is governed primarily by the expectation value of discrete rotational symmetry, which is robust against changes in microscopic parameters.

In this paper, we have considered the three-electron system in order to compare with the one-electron calculations in Sec.~\ref{sec:chiral}, but the same correlation can also be observed in the two-electron system. The results for the two-electron case are presented in Appendix~\ref{sec:hydro_n2}.

\section{Summary and discussion \label{sec:summary}}

In this study, we investigated two distinct measures of chirality in electronic systems, electron chirality $\tau^Z(\bm r) = \psi^\dg(\bm r) \dvec{\bm p} \cdot \bm \sg \psi(\bm r)$ and hydrodynamic helicity $H(\bm r) = :\bm j(\bm r) \cdot (\nabla \times \bm j(\bm r)):$, and clarified their microscopic origins.
By introducing the atomic model with chiral crystal fields, we systematically analyzed how the interplay among the microscopic parameters generates finite chiralities in electronic systems.

Electron chirality $\tau^Z(\bm r)$ is a relativistic one-body quantity and is non-zero only in the presence of both spin-orbit coupling and a chiral crystal field.
Our numerical calculations and first-order perturbative analysis show that electron chirality increases with both spin-orbit coupling strength and the magnitude of the chiral crystal field. 
However, in the vicinity of quasidegenerate energy levels, electron chirality is effectively independent of SOC, leading to a remarkable enhancement.

In contrast, hydrodynamic helicity $H(\bm r)$ is a two-body pseudoscalar quantity that can emerge even without spin-orbit coupling through electron-electron interactions. We demonstrated that this quantity grows linearly with the interaction strength and is governed by symmetry selection rules associated with $C_2$ rotational symmetry.
Unlike electron chirality $\tau^Z(\bm r)$, hydrodynamic helicity $H(\bm r)$ does not exhibit divergent enhancement near degeneracy points, which we attributed to the vanishing of relevant matrix elements enforced by $C_2$ symmetry.
Furthermore, we uncovered a strong correlation between hydrodynamic helicity and the expectation value of the $C_2$ operator, indicating its character as a chirality measure controlled by discrete symmetry.

These results provide a unified microscopic understanding of electronic chirality, distinguishing between spin-orbit-coupling-driven and correlation-driven mechanisms.
The present findings offer fundamental guidelines for designing and controlling chiral electronic phenomena, including spin-dependent transport and optical responses, and establish a basis for exploring novel chiral functional materials.

The insights obtained from the model studied in this work are expected to be applicable to electronic states in ionic crystals and molecular systems.  
In solids, the situation is non-trivial due to the presence of crystal momentum $\bm{k}$, but it may correspond to cases where the electronic states at each $\bm{k}$ vary depending on the atomic configuration within the unit cell (particularly at the $\Gamma$ point with $\bm{k}=\bm 0$).  
In any case, the microscopic origin of electron chirality in molecules and solids has not yet been fully clarified, but the findings of this study are expected to provide useful guidance for such analyses.

\begin{acknowledgments}
TM is grateful to K. Hattori for fruitful discussions.
TM thanks S. Kuniyoshi and S. Hayami for sharing unpublished work on the two-body chirality measure in the multipole framework.
It would be interesting to investigate the relationship between hydrodynamic helicity and this two-body multipole.
This work was supported by the KAKENHI Grants No.~23KJ0298 (TM), No.~23H04869 (TM), No.~23K25827 (HI, MTS, SH), No.~24K00588 (MTS), No.~24K00581 (MTS), No.~25K21684 (MTS), No.~25K00947 (MTS), No.~24K00578 (SH).
\end{acknowledgments}

\appendix

\section{Relation between electron chirality and Berry connection \label{sec:berry}}

In the absence of SOC, the electron chirality can be expressed in terms of the Berry connection matrix with respect to the atomic position $\bm R_I$. 
Below, we demonstrate this relationship explicitly.
We reconsider the Hamiltonian $\mscr H^{(0)} = \int \diff\bm r \psi^\dg(\bm r)[\bm p^2/2m + V(\bm r)]\psi(\bm r)$ where $V(\bm r) = -Ze^2/|\bm r| + \sum_{I=1}^N Q_I / |\bm r - \bm R_I| = \sum_{I=0}^N Q_I / |\bm r - \bm R_I|$ with $\bm R_0 = 0, Q_0 = -Ze^2$ includes the nuclear potential and the crystal fields.
The spin and momentum degrees of freedom are decoupled when SOC is absent, and thereby the matrix element of the chirality separates into two parts:
\begin{align}
    \braket{\al|\bm p \cdot \bm \sg | \be}_0 = \braket{\al |\bm p | \be}_0 \cdot \bm \sg. \label{eq:chirality_woSOC}
\end{align}
Here, we focus on the matrix element $\braket{\al |\bm p | \be}$.
Since the relation $[\bm p, \mcal H(\bm r)] = \bm p V(\bm r)$ is satisfied, we obtain
\begin{align}
    \braket{\al|\bm p | \be}_0
    &= -\sum_{I=0}^N \imu\hbar\frac{\braket{\al |(\nabla_{\bm R_I} V(\bm r))| \be}_0}{E_\be^{(0)} - E_\al^{(0)}}, \label{eq:pmn}
\end{align}
where we have used the fact that $V(\bm r)$ depends only on $\bm r - \bm R_I$.

To see the relation between Eq.~\eqref{eq:pmn} and the Berry connection matrix, we consider the perturbation theory \cite{Vanderbilt_book, Resta92, King-Smith1993}.
The differentiation of the Schr\"odinger equation is given as $(E_n^{(0)} - \mcal H)\ket{\nabla_{\bm R_I} \be} = \nabla_{\bm R_I} (\mcal H - E_n^{(0)}) \ket{\be}$.
This equation can also be obtained in usual perturbation theory with the perturbative Hamiltonian $\nabla_{\bm R_I} \mcal H$.
Thus, the following $\ket{\nabla_{\bm R_I} \be}_0$ satisfies the equation:
\begin{align}
    \ket{\nabla_{\bm R_I} \be}_0
    &= -\imu \bm A_\be(\bm R_I) \ket{\be} \nt
    &+ \sum_{m\in Q_0} \ket{\al} \frac{\braket{\al |(\nabla_{\bm R_I} V(\bm r))| \be}_0}{E_\be^{(0)} - E_\al^{(0)}}, \label{eq:pert_R}
\end{align}
with $\bm A_\be(\bm R) = \imu\braket{\be|\nabla_{\bm R_I} \be}_0$.
By comparing Eq.~\eqref{eq:pert_R} with Eqs.~\eqref{eq:chirality_woSOC} and \eqref{eq:pmn}, we obtain
\begin{align}
    \braket{\al|\bm p \cdot \bm \sg| \be}_0 = \hbar\sum_{I=0}^N \bm A_{\al \be}(\bm R_I) \cdot \bm \sg,
\end{align}
where $\bm A_{\al \be}(\bm R_I) = \imu\braket{\al|\nabla_{\bm R_I} \be}$ is the Berry connection matrix in terms of $\bm R_I$.

\section{Classical hydrodynamic helicity \label{sec:classical}}

We briefly review the properties of the hydrodynamic helicity in classical systems.

\subsection{
Complex classical field
}

Based on the Ginzburg-Landau (GL) theory,
the current density $\bm j$ is given by
\begin{align}
    \bm j&= - \frac{\imu \hbar q}{2m} (\psi^\ast \bm \nabla \psi - \bm \nabla \psi^\ast \psi)
    - \frac{q^2}{mc} \psi^\ast \psi \bm A,
\end{align}
where $m$ and $q$ are the mass and charge of the underlying quasiparticle.
Here $\psi$ is a complex field.
Writing $\psi = \sqrt n \epn^{\imu \theta}$, we obtain
\begin{align}
    \bm j &= \frac{\hbar q n}{m} (\bm\nabla \theta  - \frac{q}{\hbar c} \bm A)
    = q n \bm v,
\end{align}
and then
\begin{align}
    \bm j\cdot (\bm \nabla \times \bm j) &= -\frac{q^3 n^2}{mc} \bm B\cdot \bm v.
\end{align}
The total helicity is 
\begin{align}
    \int_V \frac{\bm j\cdot (\bm \nabla \times \bm j)}{n^2} \diff \bm r &= \frac{q^4}{m^2c^2} \int_V \bm A\cdot \bm B \diff \bm r,
    \label{eq:hydro_complex_field_gauge_field}
\end{align}
which is analogous to fluxoid quantization:
\begin{align}
    \oint_{\partial S} \frac{\bm j}{n} \cdot \diff \bm s
    +\frac{q^2}{mc} \int_S \bm B\cdot \diff \bm S
    &
    = \frac{\hbar q}{m} 2 \pi \mathbb Z.
\end{align}
Namely, the matter field is closely related to the electromagnetic field.

In Ref.~\cite{Hoshino2024}, we mentioned that the hydrodynamic helicity is zero for the classical field, while it takes non-zero values as shown in Eq.~\eqref{eq:hydro_complex_field_gauge_field}.
This is because the effect of the gauge field is considered in this section, but is absent in Ref.~\cite{Hoshino2024}.

\subsection{Maxwell equation}
Once $\bm j$ is given classically, we can rewrite the current density in terms of purely electromagnetic fields with the Maxwell equation:
\begin{align}
    \bm j &= \frac{c}{4\pi} ( \bm \nabla \times \bm B - \frac 1 c \dot{\bm E} ).
\end{align}
Then the hydrodynamic Helicity is given by
\begin{align}
    \int_V \bm j\cdot (\bm \nabla \times \bm j) \ \diff \bm r
    &=  \left( \frac{c}{4\pi} \right)^2 \int_V \Box \bm A \cdot \Box \bm B \ \diff \bm r,
\end{align}
where $\Box = \frac{1}{c^2}\partial_t^2 - \bm \nabla^2$ is the d'Alembertian.
The quantity on the right-hand side is reminiscent of the magnetic helicity $\int \bm A \cdot \bm B \ \diff \bm r$ \cite{Woltjer1958}.

\subsection{Linking number in stationary state}

The conservation law of charges is given by
\begin{align}
    \del_t \rho + \bm \nabla \cdot \bm j = 0. \label{eq:conservation}
\end{align}
In the stationary case, the conservation law is given by $\bm \nabla \cdot \bm j = \bm 0$.
Now, we reconsider the vorticity:
\begin{align}
    \bm \Omega = \bm \nabla \times \bm j.
\end{align}
Acting $\bm \nabla \times$ from the left-hand side and using the relation $\bm \nabla \times (\bm \nabla \times \bm j) = \bm \nabla(\bm \nabla \cdot \bm j) - \bm \nabla^2 \bm j$ and the conservation law, we have the Poisson equation:
\begin{align}
    \bm \nabla^2 \bm j = -\bm \nabla \times \bm \Omega.
\end{align}
This equation can be solved by introducing the Green's function $\bm \nabla^2 G(\bm r - \bm r') = \delta(\bm r - \bm r')$ with $4\pi G(\bm r) = -1/|\bm r|$.
Then, we obtain 
\begin{align}
    \bm j(\bm r) 
    &= \frac{1}{4\pi} \int \diff\bm r' \bm \Omega(\bm r') \times \frac{(\bm r - \bm r')}{|\bm r - \bm r'|^3}.
\end{align}
Inserting this into the definition of hydrodynamic helicity, we have 
\begin{align}
    \bm j(\bm r) \cdot [\bm \nabla \times \bm j(\bm r)] = \int \diff \bm r' \frac{[\bm \Omega(\bm r) \times \bm \Omega(\bm r')] \cdot (\bm r - \bm r')}{|\bm r - \bm r'|^3}.
\end{align}
When the vorticity $\bm \Omega(\bm r)$ forms flux lines in three-dimensional space, this expression represents a linking number of the flux lines \cite{Moffatt1992, Moffatt2014}.

In the more general perspective, we can consider the several types of linking number.
Namely, in the stationary state, we have the relations
\begin{align}
    \bm \Omega &= \bm \nabla \times \bm j
    = - \frac{c}{4\pi} \bm \nabla^2 \bm B,
    \\
    \bm j &= \frac{c}{4\pi} \bm \nabla \times \bm B
    = - \frac{c}{4\pi} \bm \nabla^2 \bm A,
    \\
    \bm B &= \bm \nabla \times \bm A.
\end{align}
Then we define the three kinds of helicities, each of which represents a linkage of flux lines:
\begin{align}
    \begin{matrix}
        \int  \bm j \cdot (\bm \nabla \times \bm j)\ \diff \bm r
         & & \text{Flux-line linkage of } \bm \Omega \\[1mm]
        \int  \bm B \cdot (\bm \nabla \times \bm B)\ \diff \bm r
         &\longleftrightarrow& \text{Flux-line linkage of } \, \bm j \,  \\[1mm]
        \int  \bm A \cdot (\bm \nabla \times \bm A)\ \diff \bm r
         & & \text{Flux-line linkage of } \bm B 
    \end{matrix}
    \nonumber
\end{align}
These are, respectively, hydrodynamic helicity \cite{Moreau1961}, (a part of) Lipkin's zilch \cite{Lipkin1964}, and magnetic helicity \cite{Woltjer1958}.
Although in general, these vector fields are continuously distributed, by considering them as bundles of closed flux lines, one can introduce the concept of the linking number.

\begin{figure}[tb]
    \centering
    \includegraphics[width=8.5cm]{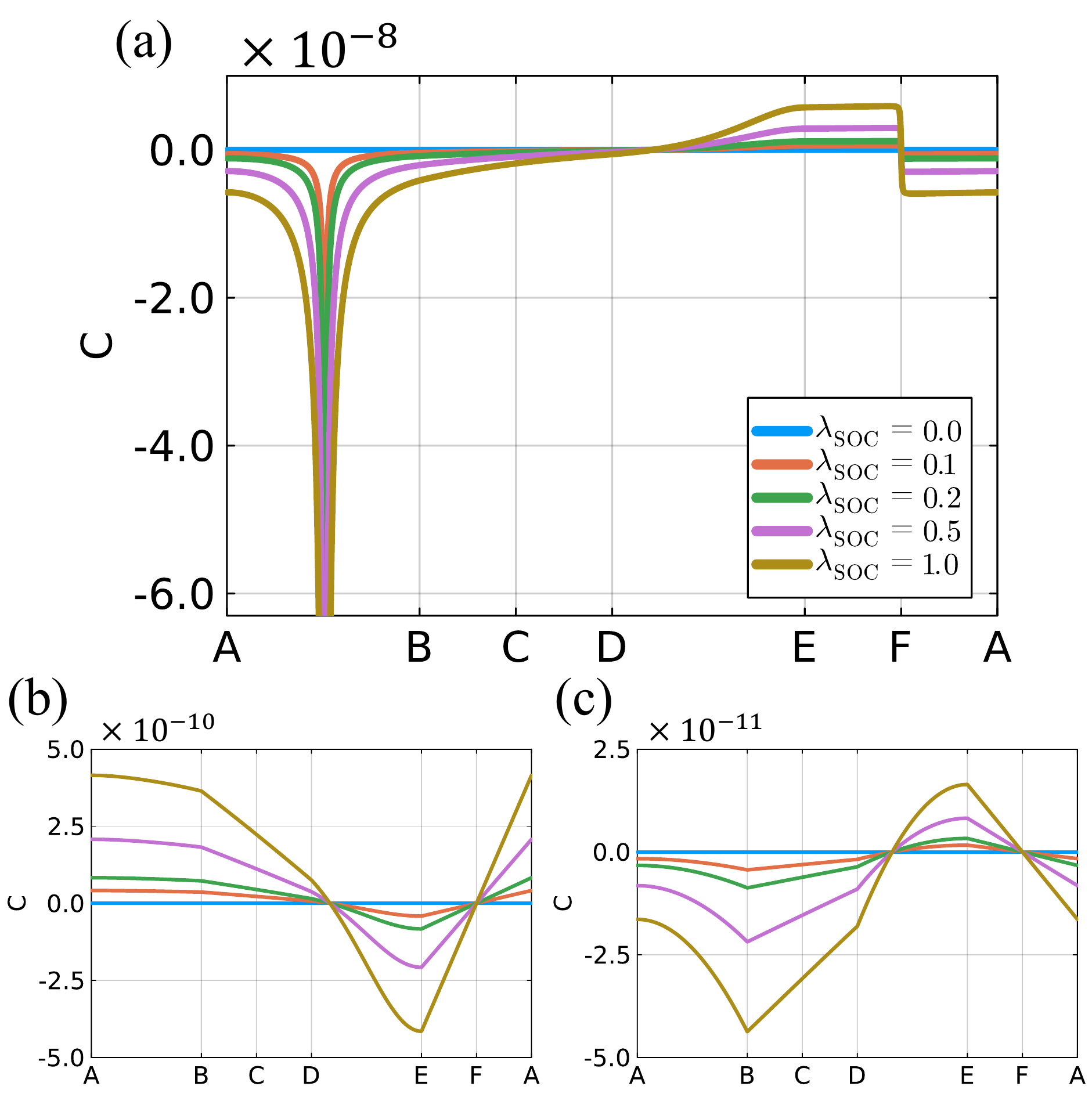}
    \caption{
    (a) Electron chirality for the second excited state.
    (b) Sum of the electron chiralities of the first and second excited states [Fig.~\ref{fig:chirality_1st} (a) $+$ Fig.~\ref{fig:chirality_2nd} (a)].
    (c) Sum of the electron chiralities of the ground, first, and second excited states [Fig.~\ref{fig:chirality_gs} $+$ Fig.~\ref{fig:chirality_1st} (a) $+$ Fig.~\ref{fig:chirality_2nd} (a)].
    }
    \label{fig:chirality_2nd}
\end{figure}

\section{Electron chirality for the second excited state \label{sec:chirality_2nd}}

In this paper, we focus on the electron chirality of the ground states and the first excited states.
As shown in Fig.~\ref{fig:energy}, $p$-like states appear in the low-energy region, and there are energy crossings between the first and second excited states. 
Here, we present the results for the electron chirality of the remaining second excited state; namely, we choose it as $\mcal P$ in Eq.~\eqref{eq:chirality_tot}.

Figure~\ref{fig:chirality_2nd} (a) shows the electron chirality of the second excited state. 
One finds a negative peak at the same position as the positive peak of the chirality for the first excited state shown in Fig.~\ref{fig:chirality_1st}(a), together with a drastic sign change at the F point.
This behavior appears to be almost identical to that of the first excited state with an opposite sign.

Figure~\ref{fig:chirality_2nd} (b) presents the sum of the chirality for the first and second excited states.
Most of the electron chirality is canceled out, leaving a finite residual of the order of $10^{-10}$.
This magnitude is comparable to that of the ground-state chirality in Fig.~\ref{fig:chirality_gs}.
Moreover, the behavior of chirality in Fig.~\ref{fig:chirality_2nd} (b) closely resembles the ground-state chirality with its sign reversed.
These results indicate that the chirality originating from $p$-like orbitals largely cancels out when all contributions are summed, which is known in the molecules \cite{Hara2012, Senami2019, Kuroda2022} and in solids \cite{Miki2025}.
Nevertheless, even after summing over all $p$-like states, the total chirality does not vanish completely and remains finite at the order of $10^{-11}$.

\section{Hydrodynamic helicity for same-orbital interaction \label{sec:hydro_sameorb}}

To clarify which interaction components give rise to the hydrodynamic helicity, we performed calculations including only the same-orbital interaction with $U = 1$.

\begin{figure}[tb]
    \centering
    \includegraphics[width=6.5cm]{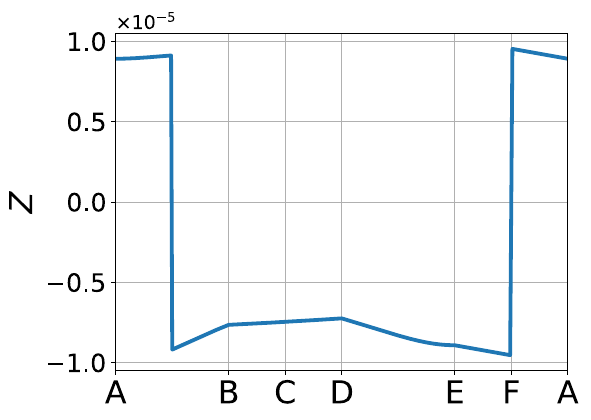}
    \caption{Hydrodynamic helicity for the same-orbital interaction.}
    \label{fig:hydro_sameorb}
\end{figure}

The results are shown in the Fig.~\ref{fig:hydro_sameorb}, and one can see that they are almost identical to those obtained when all interaction terms are included (see Fig.~\ref{fig:hydro}).
This indicates that the hydrodynamic helicity is non-zero solely due to local interaction contributions.
Furthermore, as discussed in Sec.~\ref{sec:hydro_c2}, the behavior follows that of the $C_2$ operator, which results in the universal behavior of hydrodynamic helicity.

\section{Effect of spin-orbit coupling for hydrodynaamic helicity \label{sec:hydro_soc}}

\begin{figure}[tb]
    \centering
    \includegraphics[width=6.5cm]{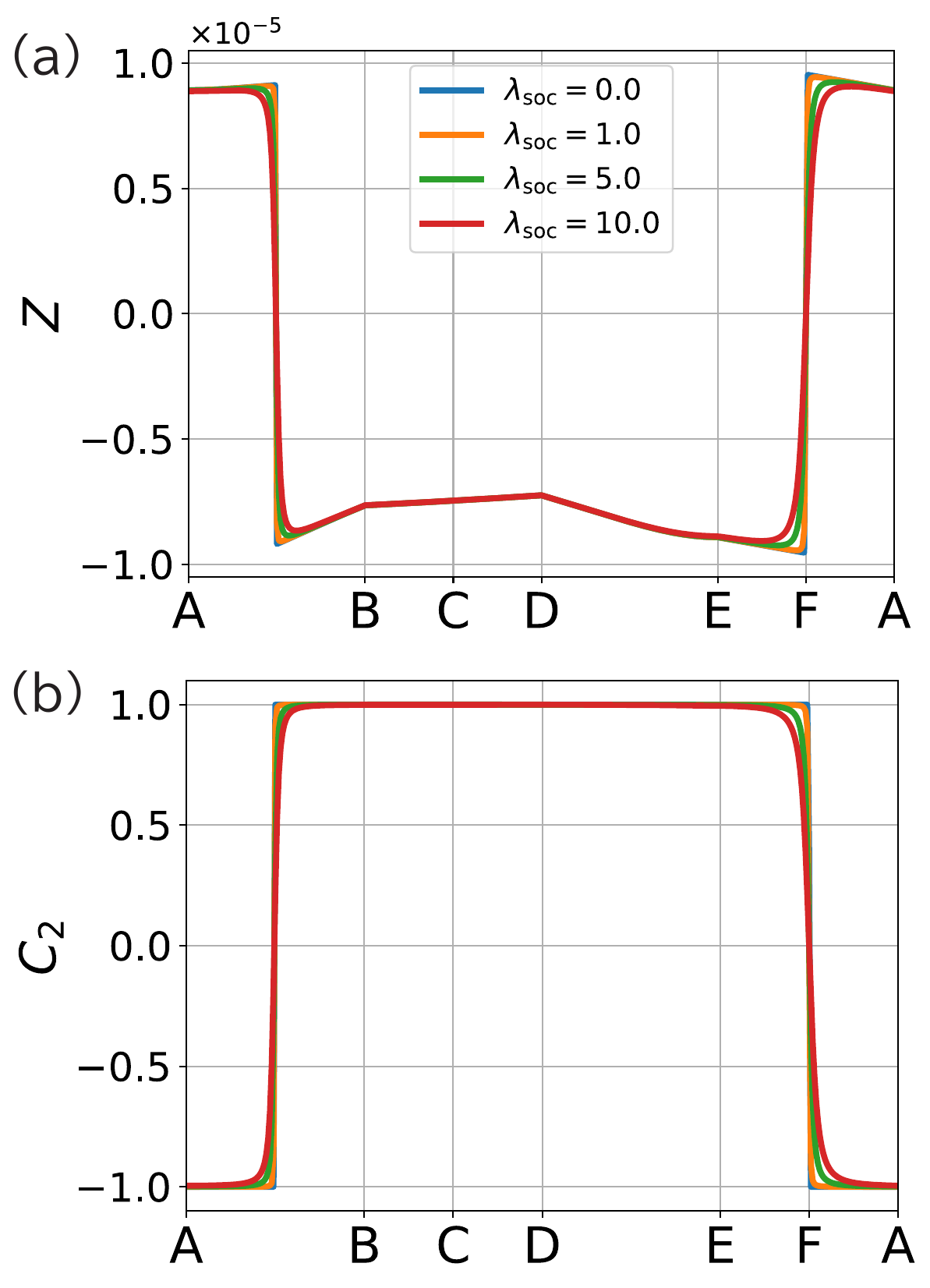}
    \caption{Effect of SOC for hydrodynamic helicity. 
    (a) hydrodynamic helicity and (b) expectation value of $C_2$ operator for several values of $\lam_{\mrm{SOC}}$.}
    \label{fig:hydro_soc}
\end{figure}

Here, we discuss the effect of SOC on the hydrodynamic helicity.
We consider the Hamiltonian $\mscr H = \mscr H_0 + \mscr H_{\mrm{CF}} + \lambda_{\mrm{SOC}}\mscr H_{\mrm{SOC}} + \mscr H_{\mrm{int}}$.
The hydrodynamic helicity for the case with SOC is shown in Fig.~\ref{fig:hydro_soc} (a).
When SOC is introduced ($\lam_{\mrm{SOC}} > 0$), the sign reversal becomes broader. 
We also show the expectation value of the $C_2$ rotational operator in Fig.~\ref{fig:hydro_soc} (b).
As in the $C_2$-symmetry breaking case in Sec.~\ref{sec:hydro_c2}, we again find a clear correlation between $Z$ in Fig.~\ref{fig:hydro_soc} (a) and the expectation value of the $C_2$ operator in Fig.~\ref{fig:hydro_soc} (b).
Therefore, the SOC has an effect on the hydrodynamic helicity similar to that of introducing $C_2$ symmetry breaking by $u$.

\section{Hydrodynamic helicity for two-electron case \label{sec:hydro_n2}}

\begin{figure*}[t]
    \centering
    \includegraphics[width=11cm]{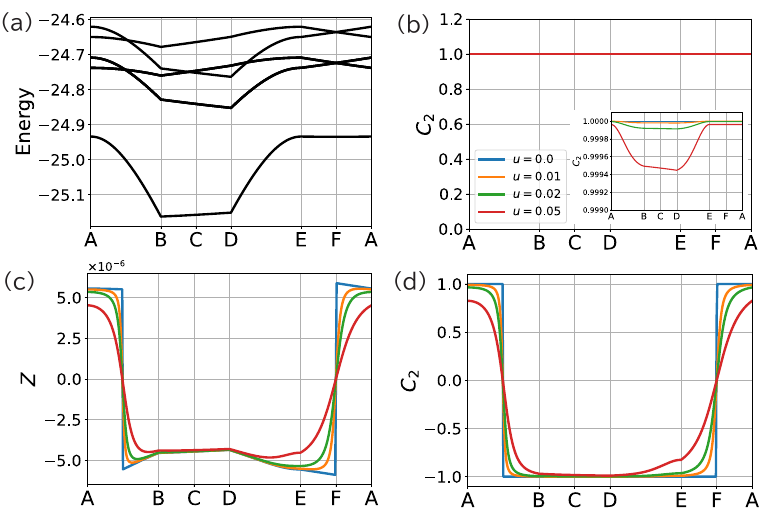}
    \caption{(a) Energy diagram for the two-electron case. 
    (b) Expectation value of $C_2$ operator for (singlet) ground state. 
    The inset shows an enlarged view around $C_2 = 1$.
    (c) Hydrodynamic helicity and (d) expectation value of $C_2$ operator for (triplet) excitation states.
    In the ground state, the hydrodynamic helicity is zero along the whole path.}
    \label{fig:hydro_n2}
\end{figure*}

We investigate the hydrodynamic helicity for the two-electron system.
Figure~\ref{fig:hydro_n2} (a) shows the energy diagram for the two-electron case.
Here, we focus on the ground state and first excited state.
The ground state is a non-degenerate (spin-singlet) state, and the first excited states are three-degenerate (spin-triplet) states.

We calculate the hydrodynamic helicity and the expectation value of the $C_2$ operator as discussed in Sec.~\ref{sec:hydro_c2}.
In the ground state, the hydrodynamic helicity is zero for any $u$, and the expectation value of the $C_2$ operator in this case is shown in Fig.~\ref{fig:hydro_n2} (b).
The expectation value of the $C_2$ operator is 1 in the $u = 0$ case, and although it deviates slightly from 1 for $u > 0$ [inset of Fig.~\ref{fig:hydro_n2} (b)], it remains almost unchanged overall.

As for the first excited state, the hydrodynamic helicity takes non-zero values as presented in Fig.~\ref{fig:hydro_n2} (c), and the corresponding expectation value of $C_2$ operator is shown in Fig.~\ref{fig:hydro_n2} (d).
In Fig.~\ref{fig:hydro_n2} (c), a sign reversal is observed at the energy crossing, in a manner similar to the three-electron case. 
As shown in Fig.~\ref{fig:hydro_n2} (d), the expectation value of $C_2$ also has a sign change at the energy crossing point. 
These results further demonstrate a clear correlation between $Z$ and the expectation value of $C_2$.

\section{Helicity in momentum space \label{sec:momentum}}

From the quantum geometric tensor, we define $\bm A(\bm k)$ (Berry connection) and $\bm B(\bm k) = \bm\nabla_{\bm k} \times \bm A$ (Berry curvature).
Then, it is tempting to consider $\int \diff \bm k\ \bm A(\bm k) \cdot \bm B(\bm k)$.
This quantity represents the linkage of the nodal line structures \cite{Moffatt1992}.
Since the configuration of nodal lines is associated with surface states \cite{Vanderbilt_book}, the helicity in momentum space is useful for understanding the nodal line structures.
The linking number for the non-Abelian gauge fields \cite{Witten1989} is also considered for the multiband systems \cite{Vanderbilt_book, Qi2008, Essin2009}.

\bibliography{hydro}
\bibliographystyle{apsrev4-2}

\end{document}